%% file: paper.tex
\documentclass[a4paper,11pt,headings=big,DIV=12]{scrartcl}
\pdfoutput=1
\usepackage{amsmath,amssymb}
\usepackage{color,xcolor}
\definecolor{blue}{rgb}{0,0,0.5}
\usepackage[colorlinks,linkcolor=blue,urlcolor=blue,citecolor=blue]{hyperref}
\usepackage{graphicx}
\addtokomafont{disposition}{\rmfamily\boldmath}
\usepackage{cite}
\usepackage{xspace}
\usepackage{relsize}
\usepackage{bm}
\usepackage{multirow}
\usepackage{booktabs}
\usepackage[utf8]{inputenc}
\usepackage{arydshln}

\makeatletter
\def\adl@drawiv#1#2#3{%
        \hskip.5\tabcolsep
        \xleaders#3{#2.5\@tempdimb #1{1}#2.5\@tempdimb}%
                #2\z@ plus1fil minus1fil\relax
        \hskip.5\tabcolsep}
\newcommand{\cdashlinelr}[1]{%
  \noalign{\vskip\aboverulesep
           \global\let\@dashdrawstore\adl@draw
           \global\let\adl@draw\adl@drawiv}
  \cdashline{#1}
  \noalign{\global\let\adl@draw\@dashdrawstore
           \vskip\belowrulesep}}
\makeatother

\usepackage{rotating}

\allowdisplaybreaks

\DeclareOldFontCommand{\bf}{\normalfont\bfseries}{\mathbf}

\newcommand{\mysmall}[1]{\scriptscriptstyle #1} 

\def \refeq#1{(\ref{#1})}
\def \refsec#1{section~\ref{#1}}
\def \refSec#1{Section~\ref{#1}}
\def \refapp#1{app.~\ref{#1}}
\def \reffig#1{figure~\ref{#1}}
\def \refFig#1{Figure~\ref{#1}}
\def \reftab#1{table~\ref{#1}}
\def \refTab#1{Table~\ref{#1}}

\newcommand{\muNP}{{\mu_{\Lambda}}}
\newcommand{\muEW}{{\mu_\mathrm{ew}}}
\newcommand{\muLow}{{\mu}}

\def\epe{\varepsilon'/\varepsilon}
\def\epsK{\varepsilon_K}
\def\kpn{K^+\rightarrow\pi^+\nu\bar\nu}
\def\klpn{K_{L}\rightarrow\pi^0\nu\bar\nu}

\def\klpll{K_L\to\pi^0\ell\bar\ell}
\def\klpee{K_L\to\pi^0 e\bar{e}}

\newcommand{\Nf}{N_f}

\newcommand{\wc}[3][{}]{\big[{\cal C}_{#2}^{#1}\big]_{#3}}
\newcommand{\wcHat}[3][{}]{\big[\widehat{\cal C}_{#2}^{\,#1}\big]_{#3}}

\newcommand{\Wc}[2][{}]{{\cal C}_{#2}^{#1}}

\newcommand{\op}[3][{}]{[{\cal O}_{#2}^{#1}]_{#3}}
\newcommand{\opHat}[3][{}]{[\widehat{\cal O}_{#2}^{#1}]_{#3}}
\newcommand{\Op}[2][{}]{{\cal O}_{#2}^{#1}}

\begin{document}

\begin{flushleft}
\end{flushleft}

\vspace{-14mm}
\begin{flushright}
  TUM-HEP-1153/18 \\
  AJB-18-7
\end{flushright}

\vspace{0mm}

\begin{center}
{\LARGE\bfseries \boldmath
\vspace*{1.5cm}
Anatomy of $\epe$ beyond the Standard Model
}\\[0.8 cm]
{\textsc{
Jason Aebischer$^a$, Christoph Bobeth$^{a,b}$, Andrzej J.\ Buras$^c$, David M.\ Straub$^a$
}\\[0.5 cm]
\small
$^a$ Excellence Cluster Universe, Boltzmannstr.~2, 85748~Garching, Germany \\
$^b$ Physik Department, TU M\"unchen, James-Franck-Stra{\ss}e, 85748 Garching, Germany \\
$^c$   TUM Institute for Advanced Study, Lichtenbergstr.~2a, 85748 Garching, Germany
}
\\[0.5 cm]
\footnotesize
E-Mail:
\texttt{jason.aebischer@tum.de},
\texttt{christoph.bobeth@tum.de},
\texttt{andrzej.buras@tum.de},
\texttt{david.straub@tum.de}
\\[0.2 cm]
\end{center}

\bigskip

\begin{abstract}\noindent
We present for the first time a model-independent anatomy of the ratio $\epe$
in the context of  the $\Delta S = 1$ effective theory with operators invariant
under QCD and QED and in the context of the Standard Model Effective Field Theory
(SMEFT) with the operators invariant under the full SM gauge group. Our goal is
to identify the new physics scenarios that are probed by this ratio
and which could help to explain a possible deviation from the SM that is hinted
by the data.
To this end we derive a master formula for $\epe$, which
can be applied to any theory beyond the Standard Model (BSM) in which the
Wilson coefficients of all contributing operators have been calculated at the
electroweak scale. The relevant hadronic matrix elements of BSM operators are
from the Dual QCD approach and the SM ones from lattice QCD.
Within SMEFT, the constraints from $K^0$ and $D^0$
mixing as well as electric dipole moments limit significantly potential new physics
contributions to $\epe$. Correlations of $\epe$ with
$K\to\pi\nu\bar\nu$ decays are briefly discussed.
Building on our EFT analysis and the model-independent constraints,
we discuss implications of a possible deviation from the SM in $\epe$
for model building,
highlighting the role of the new scalar and tensor matrix elements in models
with scalar mediators.
\end{abstract}

\newpage

\setcounter{tocdepth}{2}
\tableofcontents

\newpage

%
%
%
\section{Introduction}

One of the stars of flavour physics since the early 1980s has been the ratio $\epe$
that measures the size of direct CP violation in $K_L\to\pi\pi$ relative to the
indirect CP violation described by $\varepsilon_K$. On the experimental side, the
world average from the NA48 \cite{Batley:2002gn} and KTeV \cite{AlaviHarati:2002ye,
Abouzaid:2010ny} collaborations reads
\begin{align}
  \label{EXP}
  (\epe)_\text{exp} &
  = (16.6 \pm 2.3) \times 10^{-4} \,.
\end{align}
On the theory side, a long-standing challenge in  making predictions for $\epe$
within the Standard Model (SM) has been the significant cancellation between QCD
and electroweak penguin contributions to this ratio. In the SM, QCD
penguins give a positive contribution and electroweak penguins a negative one.
Therefore, in order to obtain an accurate prediction, both the short-distance
contributions to this ratio, represented by Wilson coefficients of penguin operators,
as well as the long-distance hadronic matrix elements of these operators have
to be accurately known.

As far as the short-distance contributions are concerned, they have been known
already for 25~years at next-to-leading order (NLO) \cite{Buras:1991jm, Buras:1992tc, Buras:1992zv,
Ciuchini:1992tj, Buras:1993dy, Ciuchini:1993vr}. First steps towards next-to-next-to-leading order (NNLO) predictions
for $\epe$  have been made in \cite{Bobeth:1999mk, Buras:1999st, Gorbahn:2004my,
Brod:2010mj} and further progress towards a complete NNLO result is under
way~\cite{Cerda-Sevilla:2016yzo}.

The situation with hadronic matrix elements is another story and even if
significant progress on their evaluation has been made  over the last 25 years,
the present status is far from satisfactory. The situation of
$\epe$ in the SM can be briefly summarized as follows:

\begin{itemize}
\item
The analysis of $\epe$ by the RBC-UKQCD collaboration based on their lattice QCD
calculation of $K\to \pi\pi$ matrix elements \cite{Bai:2015nea, Blum:2015ywa}, as well as the analyses performed in  \cite{Buras:2015yba,Kitahara:2016nld}
that are based on the same
matrix elements but also include isospin breaking effects,
find $\epe$ in the ballpark of $(1-2) \times 10^{-4}$. This is
by one order of magnitude below the data, but with an error in the ballpark of
$5\times 10^{-4}$. Consequently, based on these analyses, one can talk about an $\epe$ anomaly of at most~$3\sigma$.
\item
An independent analysis based on hadronic matrix elements from the Dual QCD (DQCD)
approach \cite{Buras:2015xba, Buras:2016fys} gives a strong support to these values
and moreover provides an \textit{upper bound} on $\epe$ in the ballpark of $6\times
10^{-4}$.
\item
A different view has been expressed in \cite{Gisbert:2017vvj} where, using ideas
from chiral perturbation theory, the authors find $\epe = (15 \pm 7) \times 10^{-4}$.
While in
agreement with the measurement, the large uncertainty, that expresses
the difficulties in matching long distance and short distance contributions
in this framework, does not allow for clear-cut conclusions.
Consequently, values above $2\times 10^{-3}$, that are rather unrealistic from the
point of view of lattice QCD and DQCD, are not excluded in this approach.
\end{itemize}

Here, we would like to point out that all the existing estimates of $\epe$ at
NLO suffer from unaccounted-for short-distance renormalization scheme uncertainties
in the electroweak penguin contributions that are removed in the NNLO matching at
the electroweak scale \cite{Buras:1999st}. In the naive dimensional regularization
(NDR) scheme, used in all recent analyses, these corrections enhance parts of the
electroweak penguin contribution by roughly $16\%$, thereby leading to a {\em negative}
shift of $-1.3\times 10^{-4}$ decreasing the value of $\epe$, similarly to
isospin breaking effects. This could appear small in
view of other uncertainties. However, on the one hand, potential scale and
renormalization scheme uncertainties have been removed in this manner and on
the other hand, one day such corrections could turn out to be relevant. Finally,
the fact that this correction further decreases $\epe$ within the SM gives
another motivation for the search for new physics responsible
for it, and thus for the present analysis.

Based on the results from RBC-UKQCD and the DQCD approach of 2015 and
without the inclusion of NNLO corrections mentioned above, a number of
analyses have been performed in specific models beyond the SM (BSM)
with the goal to obtain a  sufficient upward shift in $\epe$ and thereby
its  experimental value. These include in
particular tree-level $Z^\prime$ exchanges with explicit realization in 331
models \cite{Buras:2015kwd, Buras:2016dxz} or models with tree-level $Z^0$
exchanges \cite{Bobeth:2017xry, Endo:2016tnu} with explicit realization in
models with mixing of heavy vector-like fermions with ordinary fermions~\cite{Bobeth:2016llm}
and the Littlest Higgs model with T-parity
\cite{Blanke:2015wba}. Also simplified $Z^\prime$ scenarios\cite{Buras:2015yca,
Buras:2015jaq}, the MSSM \cite{Tanimoto:2016yfy, Kitahara:2016otd, Endo:2016aws,
Crivellin:2017gks, Endo:2017ums}, the type-III Two-Higgs Doublet model
(2HDM) \cite{Chen:2018ytc, Chen:2018vog}, a $SU(2)_L \otimes SU(2)_R \otimes U(1)_{B-L}$
model \cite{Haba:2018byj, Haba:2018rzf} and the one based on SU(8) symmetry \cite{Matsuzaki:2018jui}
are of help here. On the other
hand, as demonstrated in \cite{Bobeth:2017ecx}, it is very unlikely that
leptoquarks are responsible for the $\epe$ anomaly when the constraints from
rare semi-leptonic and leptonic~$K$ decays are taken into account.

An important limitation of the recent literature is that it addressed the $\epe$
anomaly only in models in which new physics (NP) entered exclusively through
modifications of the Wilson coefficients of SM operators. However, generally,
BSM operators with different Dirac structures -- like the ones resulting from
tree-level scalar exchanges and leading to scalar and tensor operators -- or
chromo-magnetic dipole operators could play a significant role in $\epe$.
Until recently, no quantitative judgment of the importance of such operators
was possible because of the absence of even approximate calculations of the
relevant hadronic matrix elements in QCD.
This situation has been changed through the calculation of the matrix elements
in question for the chromo-magnetic dipole operators by lattice QCD
\cite{Constantinou:2017sgv} and DQCD \cite{Buras:2018evv} and in particular
through the calculation of matrix elements of all four-quark BSM operators,
including scalar and tensor operators, by DQCD \cite{Aebischer:2018rrz}. The
first application of these new results for chromo-magnetic dipole operators
can be found in \cite{Chen:2018vog} and in the present paper we will have a
closer look at all BSM operators.

Another important question is which of the operators in the
low-energy effective theory can be generated in a short-distance BSM scenario.
A powerful tool for this purpose is the Standard Model Effective Field
Theory (SMEFT) \cite{Buchmuller:1985jz, Grzadkowski:2010es}, where the SM
Lagrangian above the electroweak scale $\muEW \sim 100$~GeV and below the scale
of new physics $\muNP\gg \muEW$ is supplemented by all dimension five and six
operators that are invariant
under the SM gauge group $G_\text{SM} = SU(3)_c \otimes SU(2)_L \otimes U(1)_Y$.
As we will show, matching the SMEFT at tree level on the $\Delta S = 1$ effective
field theory (EFT) at $\muEW$, not all operators that are allowed by the
QCD and QED gauge symmetry $SU(3)_c \otimes U(1)_Q$ are generated.

The goal of the present paper is to perform a general BSM analysis of $\epe$,
taking into account all possible operators and exploiting the SMEFT
to single out the operators that can be generated in high-scale BSM scenarios.
In this manner, one can obtain a general view on possible BSM physics behind the
emerging $\epe$ anomaly and point out promising directions to be explored
in concrete models and exclude those in which the explanation of the data
in (\ref{EXP}) is unlikely. In the context of SMEFT, constraints from
other processes, in particular from $\epsK$, $D^0$-$\bar D^0$ mixing, and electric
dipole moments, play an important role and we will discuss them in the present paper.

One of the highlights of our paper is the derivation of a master formula for  $\epe$,
recently presented in \cite{Aebischer:2018quc}, which can be applied to any theory
beyond the SM in which the Wilson coefficients of the operators have
been calculated at the electroweak scale. The relevant hadronic matrix elements
of BSM operators entering this formula are taken from the DQCD approach and for the
SM ones from lattice QCD.

The outline of our paper is as follows. In \refsec{sec:epesm} we present a complete model-independent anatomy of $\epe$ from the point of view
of the $\Delta S = 1$ EFT and provide the master formula of $\epe$ beyond the SM.
We give also the tree-level matching of SMEFT on the $\Delta S = 1$ EFT relevant
for $\epe$. In \refsec{sec:constraints} we discuss
correlations that arise in SMEFT between $\epe$ and other processes, in particular
$\epsK$, $D^0$-$\bar D^0$ mixing, the electric dipole moment of the neutron,
and the decays $\klpn$ and $\kpn$. Based on the previous section, we derive lessons
for model building in \refsec{sec:Lessons} to facilitate
the identification of classes of models that are constrained by $\epe$ as well
as singling out prime candidates for new physics scenarios behind the $\epe$
anomaly. We summarize the main virtues of our analysis in \refsec{sec:Summary}.
In several appendices we collect our conventions, recall useful definitions,
and provide the necessary material for the numerical analysis of $\epe$ beyond the SM.

%
%
%
\section{Model-independent anatomy of $\epe$}
\label{sec:epesm}

The parameter $\epe$ measures the ratio of direct over indirect CP violation in
$K_L\to \pi\pi$ decays. Using the precisely measured $\epsK$ from experiment and
neglecting isospin breaking corrections,\footnote{%
Isospin breaking corrections have been considered in \cite{Cirigliano:2003gt,
Cirigliano:2003nn} and have been taken into account in the SM analyses
in  \cite{Buras:2015yba,Kitahara:2016nld}. There they play a significant role
in suppressing the $\text{Im}A_0$ contribution relatively to the $\text{Im}A_2$
one, making the cancellation between these two contributions stronger. However,
in BSM scenarios, such a strong cancellation
is not expected and typically contributions to  $\text{Im}A_2$ dominate as they are
not suppressed by the factor $1/\omega\approx 22$ in contrast to  $\text{Im}A_0$.
Therefore, the inclusion of isospin breaking effects in the BSM contributions
calculated by us is insignificant and it is justified to neglect them in view
of the remaining uncertainties in hadronic matrix elements that affect the
dominant contributions to $\text{Im}A_2$.
} it can be written as\footnote{%
It is common to omit the subscript $K$ on $\varepsilon\equiv\varepsilon_K$
when writing the ratio $\epe$.
}
\begin{align}
  \label{eq:epe-formula}
  \frac{\varepsilon'}{\varepsilon} &
  = -\frac{\omega}{\sqrt{2}|\epsK|}
    \left[ \frac{\text{Im}A_0}{\text{Re}A_0}
         - \frac{\text{Im}A_2}{\text{Re}A_2} \right]\,,
\end{align}
where $A_{0,2}$ are the $K\to\pi\pi$ isospin amplitudes
\begin{align}
  A_{0,2} &
  = \Big\langle (\pi\pi)_{I=0,2}\, \Big|\; \mathcal{H}_{\Delta S = 1}^{(3)}(\muLow)
  \;\Big|\, K \Big\rangle \,,
  \label{eq:A02}
\end{align}
and the ratio $\omega = {\text{Re}A_2}/{\text{Re}A_0} \approx 1/22$ expresses
the enhancement of ${\text{Re}A_0}$ over ${\text{Re}A_2}$ known
as the $\Delta I = 1/2$ rule.
$\mathcal{H}_{\Delta S = 1}^{(3)}$ denotes the effective Hamiltonian of the
$\Delta S = 1$ EFT taken at the low-energy scale $\muLow \sim 1$~GeV with only the
three lightest quarks, $q=u,d,s$ being dynamical. It is obtained by decoupling the heavy
$W^\pm$, $Z^0$, and $h^0$ bosons and the top quark at the electroweak scale $\muEW$
and the bottom and charm quarks at their respective mass thresholds $\mu_b$ and $\mu_c$,
respectively \cite{Buchalla:1995vs}.

\begin{figure}[tbp]
\centering
\includegraphics[width=\textwidth]{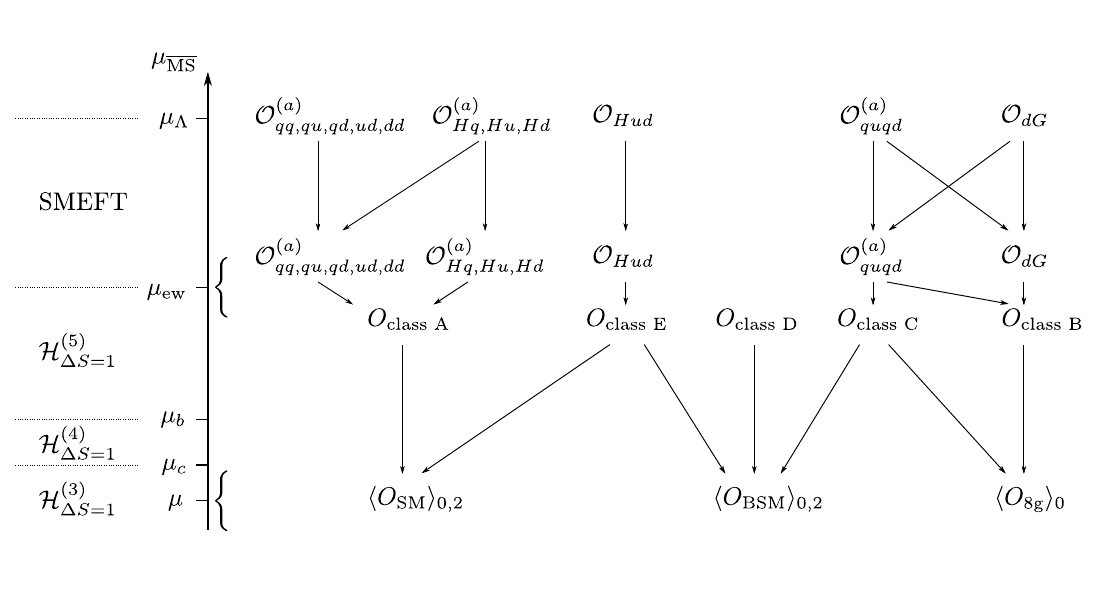}
\caption{Sketch of the different contributions to $\epe$ discussed in
  \refsec{sec:epesm}, starting from Wilson coefficients of SMEFT operators
  at a high scale $\muNP$, evolved to the electroweak scale $\muEW$ with the
  SMEFT RG equations, matched onto the 5-flavour $\Delta S=1$
  effective Hamiltonian (\refsec{sec:matching}), evolved to the hadronic scale
  $\muLow$ (\refsec{sec:rge}), and multiplied by the $K\to\pi\pi$ matrix elements
  (\refsec{sec:me}).
  In the SMEFT running, the arrows indicate operator mixing arising from top-quark
  Yukawa or gauge couplings. The matching is performed at tree level. We have
  omitted semi-leptonic and electro-magnetic dipole operators. }
\label{fig:sketch}
\end{figure}

The values of the Wilson coefficients in this effective Hamiltonian encode all
possible NP effects in $\epe$.
However, when considering a NP  model at a scale $\muNP$, much larger than the electroweak scale $\muEW$,
these low-energy Wilson coefficients are only the final step in a series of
effective theories.
At $\muNP\gg\muEW$, integrating out the heavy new particles leads to the SMEFT Lagrangian
with dimension five and six operators invariant under the full SM gauge group.
Using the SMEFT renormalization group (RG) equations, these can be evolved to $\muEW$
and matched onto $\mathcal{H}_{\Delta S = 1}^{(5)}$ with five active quark
flavours. This hierarchy of effective theories is sketched in \reffig{fig:sketch}
and the remainder of this section will be devoted to discussing the individual
steps in detail, starting from the lowest scale:
\begin{itemize}
\item \refSec{sec:me} discusses the relevant operators in
  $\mathcal{H}_{\Delta S = 1}^{(3)}$ and their $K\to \pi\pi$ matrix elements.
\item \refSec{sec:rge} discusses the RG evolution between lowest scale $\muLow$
  and $\muEW$ and the additional operators in $\mathcal{H}_{\Delta S = 1}^{(5)}$
  that can play a role.
\item \refSec{sec:master} summarizes the results of \refsec{sec:me} and
  \refsec{sec:rge} in the form of a convenient \textit{master formula} of
  $\epe$.
\item \refSec{sec:matching} discusses the matching of SMEFT onto
  $\mathcal{H}_{\Delta S = 1}^{(5)}$ at $\muEW$, singling out the operators that arise
  at the dimension-six level, and briefly discusses RG effects in SMEFT above $\muEW$.
\end{itemize}
\refFig{fig:sketch} can serve as a map guiding through this anatomy
and already anticipates some of the findings of this section.

%
%
\subsection{$K\to \pi\pi$ matrix elements}\label{sec:me}

Given the values of the Wilson coefficients in the effective Hamiltonian
\begin{align}
  \label{eq:DS1-Hamiltonian}
  \mathcal{H}_{\Delta S = 1}^{(3)} &
  = - \sum_i C_i(\muLow) \, O_i\,,
\end{align}
at the low-energy scale $\muLow$, the $K\to\pi\pi$ isospin amplitudes can be calculated
by means of \refeq{eq:A02} if the matrix elements
\begin{align}
  \label{eq:def-Kpipi-ME}
  \langle O_i (\muLow) \rangle_I &
  \equiv \left\langle (\pi\pi)_I | O_i | K \right\rangle(\muLow)\,,
\end{align}
are known at the scale $\muLow$. Neglecting electro-magnetic corrections,
only chromo-magnetic dipole operators
\begin{align}
  \label{eq:DS1-dipole-QCD}
  O_{8g}^{(\prime)}      &
  = m_s(\bar s \, \sigma^{\mu\nu} T^A P_{L(R)} d) \, G^A_{\mu\nu} \,,
\end{align}
and four-quark operators
\begin{align}
  \label{eq:DS1-psi4}
  O_{XAB}^q &
  = (\bar s^i \Gamma_X P_A d^i) (\bar q^j \Gamma_X P_B q^j) \,,
&
  \widetilde{O}_{XAB}^q &
  = (\bar s^i \Gamma_X P_A d^j) (\bar q^j \Gamma_X P_B q^i) \,,
\end{align}
can contribute.
Here $i,j$ are colour indices, $A,B=L,R$, and $X=S,V,T$ with $\Gamma_S=1$,
$\Gamma_V=\gamma^\mu$, $\Gamma_T=\sigma^{\mu\nu}$.\footnote{For $\Gamma_T$
there is only $P_A = P_B$ in four dimensions but not $P_A \neq P_B$.}
Throughout it is sufficient to consider the case $A = L$, whereas results
for the \textit{chirality-flipped} case $A = R$ (obtained by interchange of
$L \leftrightarrow R$ for both $A,B$) follow analogously due to parity
conservation of QCD and QED: the $K\to\pi\pi$ matrix elements of chirality-flipped
operators have just opposite sign.

Since the number of active quark flavours is $N_f=3$, in principle the four-quark
operators with $q=u,d,s$ are present in \refeq{eq:DS1-psi4}.
However, we expect the contribution to $K\to \pi\pi$ matrix elements from operators
with flavour structure $(\bar sd) (\bar ss)$ to be strongly suppressed\footnote{%
The $\Nf = 3$ lattice results \cite{Bai:2015nea, Blum:2015ywa} of the $K\to\pi\pi$
matrix elements in principle include these contributions but from these results
they cannot be disentangled from the $(\bar sd) (\bar uu)$ and $(\bar sd) (\bar dd)$
ones. In this regard it is desirable
that lattice collaborations provide in the future separately the matrix
element for each $(\bar sd)(\bar qq)$ operator for $q = u, d, s$.
} and we will neglect them.

Using Fierz relations to eliminate redundant operators
(see \refapp{app:fierz} for details),
it then follows that there are only $10 + 10'$ $(\bar sd)(\bar uu)$
and $5 + 5'$ $(\bar sd)(\bar dd)$ linearly independent four-quark operators
that contribute to $\epe$ via a non-vanishing $K\to \pi\pi$
matrix element and in addition the chromo-magnetic dipole operators ($1 +1'$).
In the amplitude $A_0$, there are then in total 16 independent matrix elements,
seven of which are the ones of the SM four-quark operators and one the chromo-magnetic
dipole matrix element. In the amplitude $A_2$, further simplifications arise as the
chromo-magnetic dipole operator cannot generate a $\Delta I=3/2$ transition,
neither can an operator of the form ${O}_{XAB}^u +  {O}_{XAB}^d$, leaving only
five linearly independent matrix elements, three of which are present in the SM.
We write the number of total matrix elements in the $I=0,2$ amplitudes as $16_0 + 5_2$.
In \refapp{app:Kpipi-ME}, we specify a non-redundant basis for them.

By now these matrix elements are known with varying accuracy:
\begin{itemize}
\item
First lattice calculations for the $7_0 + 3_2$ matrix elements\footnote{%
Note that the 10 operators in the traditional SM basis
are not linearly independent and correspond
only to $7$ linearly independent operators for $\Nf = 3$ \cite{Buras:1993dy}.
}
generated in the SM
have recently been performed by the RBC-UKQCD collaboration
\cite{Blum:2015ywa, Bai:2015nea}. These results are in good
agreement with the pattern of matrix elements
of the relevant QCD and QED penguin operators obtained in the DQCD
approach \cite{Bardeen:1986vz, Buras:1985yx, Buras:2014maa, Buras:2015xba}.
\item
The $K\to\pi\pi$ matrix element of the chromo-magnetic dipole operator is
presently not accessible directly on the lattice, but can only be estimated by
relating it to the analogous $K\to\pi$ matrix element via $SU(3)$ chiral symmetry \cite{Constantinou:2017sgv}.
Recently, the $K\to\pi\pi$  matrix element of this operator has been calculated
directly for the first time in the DQCD approach in the $SU(3)$ chiral limit
\cite{Buras:2018evv}. Both results are in good agreement with each other and
show that the relevant matrix element is by a factor of three to four smaller
than previously expected in the chiral quark model \cite{Bertolini:1994qk},
thereby decreasing the impact of these operators on $\epe$.
Nevertheless there are NP scenarios where they play an important role (see e.g.~\cite{Gedalia:2009ws,Chen:2018vog}).
\item
The matrix elements of the remaining $8_0+2_2$ linearly independent BSM four-quark matrix elements in \reftab{tab:me-values}
have only been calculated very recently in DQCD in the $SU(3)$ chiral limit  \cite{Aebischer:2018rrz} and it will still take some time before corresponding results in lattice QCD will be available. Yet already these approximate results from DQCD can teach us a lot about the relevance of various operators. The scalar and tensor operators $X=S,T$ belong to this group and their matrix elements cannot be expressed in terms of the SM ones.
\end{itemize}
We give the numerical values of all matrix elements in \refapp{app:Kpipi-ME}.

To summarize, there are three classes of matrix elements that
can play a role in $\epe$,
\begin{itemize}
  \item the matrix elements present in the SM,
  \item the chromo-magnetic dipole matrix element,
  \item the matrix elements of BSM scalar and tensor operators.
\end{itemize}
These three classes are indicated at the bottom of the sketch in \reffig{fig:sketch}.

%
\subsection{Renormalization group evolution below the electroweak scale}
\label{sec:rge}

In the previous subsection, we have seen that $15+15'$ four-quark operators in
$\mathcal H_{\Delta S=1}^{(3)}$ can contribute to $K\to \pi\pi$ at the scale
$\muLow$. However, additional four-quark operators
are present in the five-flavour Hamiltonian $\mathcal H_{\Delta S=1}^{(5)}$ at
$\muEW$, namely the four-quark operators with flavour structures $(\bar s d)(\bar q q)$
where $q=c$ and $b$. They can contribute to $\epe$ \textit{indirectly} if they
undergo QCD and/or QED RG mixing with $q = u, d$ operators. The same is true for
the operators with $q=s$ that were already present for $\Nf=3$, but did not contribute
\textit{directly} (at least in our approximation).
In principle, also semi-leptonic operators can contribute, since they mix
under QED into four-quark operators, but we will neglect them in the following,
since they are typically strongly constrained from semi-leptonic kaon decays
(as demonstrated for leptoquark models in \cite{Bobeth:2017ecx}).

To evolve the Wilson coefficients from $\muEW$ down to the scale $\muLow$ where
the matrix elements are evaluated, the anomalous dimension matrices (ADMs) are
required. The QCD and QED one-loop ADMs for the linearly independent set of four-quark
and dipole operators can be extracted from the literature \cite{Buras:2000if,
Aebischer:2017gaw, Jenkins:2017dyc} and we have implemented them in the open source
tool \texttt{wilson} \cite{Aebischer:2018bkb} that allows to solve the RG equations
numerically.

Inspection of the RG mixing reveals that
\begin{itemize}
\item The vector operators $O^{c,b}_{VAB}$ and their colour-flipped counterparts,
  as well as the operators $O^{s,d}_{SAB}$ with $A\neq B$,
  mix into $O^{u,d}_{VAB}$ at one loop in QCD and QED, specifically into
  the QCD and QED penguin operators present in the SM.
\item For scalar and tensor operators $O^{q}_{XAA}$
  ($X=S$ or $T$) there is instead no mixing among operators with different $q$. This
  implies in particular that the operators $O^{s,c,b}_{XAA}$ cannot mix into four-quark
  operators that have non-vanishing $K\to\pi\pi$ matrix elements. However, they
  {\em do} mix at one loop in QCD into the chromo-magnetic dipole operators
  $O_{8g}^{(\prime)}$ and in QED into the electro-magnetic ones.
\end{itemize}

Taking these observations into account, we can identify for a given chirality
five qualitatively different classes of NP models where different $K\to\pi\pi$
matrix elements are relevant:

\paragraph{Class A} Models with NP represented by the operators
\begin{equation}
  \label{A1}
  O_{VLL}^q, \qquad \widetilde{O}_{VLL}^q, \qquad
  O_{VLR}^q, \qquad \widetilde{O}_{VLR}^q ,\qquad (q=u,c,b)
\end{equation}
and
\begin{equation}
  \label{A2}
  O_{VLL}^q, \qquad  O_{VLR}^q, \qquad O_{SLR}^q, \qquad (q=d,s)
\end{equation}
as well as their chirality-flipped counterparts
at the electroweak scale contribute to $\epe$ via operators whose
matrix elements can be written as linear combinations of the matrix elements
of SM operators that were calculated in lattice QCD. Note that operators with
$q = s,c,b$ contribute via RG mixing into operators with $q = u, d$ and only
the matrix elements of the latter are related to matrix elements of the SM
operators. Therefore NP contributions in this class of models presently rely
on lattice QCD calculations \cite{Bai:2015nea, Blum:2015ywa}, which are supported
by DQCD results.

\paragraph{Class B} Models with NP represented by the operators
\begin{equation}
  \label{B1}
  O_{SLL}^{c,b}, \qquad \widetilde{O}_{SLL}^{c,b}, \qquad
  O_{TLL}^{c,b}, \qquad \widetilde{O}_{TLL}^{c,b}
\end{equation}
and
\begin{equation}
  \label{B2}
  O_{8g}, \qquad O_{SLL}^{s}, \qquad  O_{TLL}^{s},
\end{equation}
as well as their chirality-flipped counterparts
only contribute to $\epe$ through RG mixing into the chromo-magnetic
dipole operators.\footnote{As stated above, we neglect electro-magnetic
dipole operators.} The relevant matrix element has been calculated recently
by lattice QCD \cite{Constantinou:2017sgv} and DQCD \cite{Buras:2018evv}.

\paragraph{Class C}
Models with NP represented by the operators with the flavour structure
$(\bar sd)(\bar u u)$
\begin{equation}
  \label{C1}
  O_{SLL}^{u}, \qquad \widetilde{O}_{SLL}^{u}, \qquad
  O_{TLL}^{u}, \qquad \widetilde{O}_{TLL}^{u}
\end{equation}
as well as their chirality-flipped counterparts
contribute via BSM matrix elements \cite{Aebischer:2018rrz} or the chromo-magnetic
dipole matrix elements  \cite{Constantinou:2017sgv, Buras:2018evv}. None
of these matrix elements can be expressed in terms of the ones of SM  four-quark operators.

\paragraph{Class D}
Models with NP represented by the operators with the flavour structure
$(\bar sd)(\bar dd)$
\begin{equation}
  \label{D1}
  O_{SLL}^d, \qquad O_{TLL}^{d},
\end{equation}
as well as their chirality-flipped counterparts
contribute via BSM matrix elements \cite{Aebischer:2018rrz} or the chromo-magnetic
dipole matrix element \cite{Constantinou:2017sgv, Buras:2018evv}.

\paragraph{Class E}
Models with NP represented by the operators with the flavour structure
$(\bar sd)(\bar u u)$
\begin{equation}
  \label{E1}
  O_{SLR}^u, \qquad \widetilde{O}_{SLR}^{u},
\end{equation}
as well as their chirality-flipped counterparts
contribute exclusively via BSM matrix elements~\cite{Aebischer:2018rrz}
to the $I=0$ amplitude. The $I=2$ matrix elements can instead be expressed
in terms of the SM ones.

There are $37 + 37'$ operators in Classes A--E.
The only remaining $4+4'$ operators in $\mathcal H_{\Delta S=1}^{(5)}$,
namely
\begin{equation}
  O_{SLR}^{c,b}, \qquad \widetilde{O}_{SLR}^{c,b},
\end{equation}
and their chirality-flipped counterparts, have been omitted
since they neither contribute directly nor via RG mixing to $\epe$ at the level
considered.

%
\subsection{Master formula for $\epe$ beyond the SM}\label{sec:master}

Having both the RG evolution and all matrix elements at the low-energy scale
$\muLow$ for the first time at hand allowed us recently \cite{Aebischer:2018quc}
to present in a letter a master formula for  $(\epe)_\text{BSM}$ that exhibits
its dependence on each Wilson coefficient at the scale $\muEW$ and consequently
is valid in {\em any} theory beyond the SM that is free from non-standard light degrees
of freedom below the electroweak scale. We will now discuss various ingredients and
technical details which led to this formula.

The numerical analysis in  \cite{Aebischer:2018quc} has been performed with the
public codes \texttt{flavio}~\cite{flavio, Straub:2018kue} and \texttt{wilson}~\cite{Aebischer:2018bkb}.
In the evaluation of $\epe$ we set Re\,$A_{0,2}$ in \refeq{eq:epe-formula} to the
measured values \cite{Cirigliano:2011ny}
\begin{align}
\text{Re}\,A_0 &= 27.04(1) \times 10^{-8} \,\text{GeV}  \,,&
\text{Re}\,A_2 &= 1.210(2) \times 10^{-8} \,\text{GeV} \,,
\end{align}
accounting thus for potential new physics. We use here the same convention for the
normalization ($h=1$) of the amplitudes as has been chosen for the calculation of
the matrix elements of BSM operators in \cite{Buras:2018evv, Aebischer:2018rrz},
which differs from the one ($h=\sqrt{3/2}$) used by RBC-UKQCD \cite{Bai:2015nea,
Blum:2015ywa}.
We fix $\muEW = 160$~GeV, close to the top-quark mass, and $\muLow = 1.3$~GeV.
Writing $\epe$ as a sum of the SM and BSM contributions,
\begin{align}
  \frac{\varepsilon'}{\varepsilon} &
  = \left(\frac{\varepsilon'}{\varepsilon}\right)_\text{SM}
  + \left(\frac{\varepsilon'}{\varepsilon}\right)_\text{BSM} \,,
\end{align}
the master formula of \cite{Aebischer:2018quc} for the BSM part then reads\footnote{
Note that here we have the convention of dimensionful Wilson coefficients in
\refeq{eq:DS1-Hamiltonian} and \refeq{eq:DS1-Hamiltonian-5} in contrast to
\cite{Aebischer:2018quc}. Both are related by simple rescaling $C_i^\text{here} =
C_i^{}/(1\,\text{TeV})^2$, which is taken care of in \refeq{eq:master},
such that $P_i$ and $p_{ij}^{(I)}$ are the same. }
\begin{align}
  \label{eq:master}
  \left(\frac{\varepsilon'}{\varepsilon}\right)_\text{BSM} &
  = \sum_i  P_i(\muEW) ~\text{Im}\left[ C_i(\muEW) - C^\prime_i(\muEW)\right]
  \times (1\,\text{TeV})^2,
\end{align}
where
\begin{align}
  \label{eq:master2}
  P_i(\muEW) & = \sum_{j} \sum_{I=0,2} p_{ij}^{(I)}(\muEW, \muLow)
  \,\left[\frac{\langle O_j (\muLow) \rangle_I}{\text{GeV}^3}\right],
\end{align}
with the sum over $i$ extending over the Wilson coefficients $C_i$ of all
operators in Classes A--E and their chirality-flipped counterparts, that is $36 + 36'$
linearly independent four-quark operators and $1 + 1'$ chromo-magnetic dipole operators.
The $C_i'$ are the Wilson coefficients of the corresponding
chirality-flipped operators obtained by interchanging $P_L\leftrightarrow P_R$.  The
relative minus sign accounts for the fact that their $K\to\pi\pi$ matrix
elements differ by a sign.  Among the contributing operators are also operators
present already in the SM but their Wilson coefficients in \refeq{eq:master}
include only BSM contributions.

The dimensionless coefficients $p_{ij}^{(I)}(\muEW,\muLow)$ include the QCD and
QED RG evolution from $\muEW$ to $\muLow$ for each
Wilson coefficient as well as the relative suppression of the contributions to
the $I=0$ amplitude due to ${\text{Re}A_2} / {\text{Re}A_0}\ll 1$ for the matrix
elements $\langle O_j (\muLow) \rangle_I$ of all the operators $O_j$ present at
the low-energy scale, see \refapp{app:Kpipi-ME}. The index $j$ includes also $i$ so that the effect of
self-mixing is included. The $P_i(\muEW)$ do not depend on $\muLow$ to the
considered order, because the $\muLow$-dependence cancels between matrix
elements and the RG evolution operator.  Moreover, it should be emphasized that
their values are {\em model-independent} and depend only on the SM dynamics
below the electroweak scale, which includes short distance contributions down to
$\muLow$ and the long distance contributions represented by the hadronic matrix
elements. The BSM dependence enters our master formula in (\ref{eq:master}) {\em
  only} through the Wilson coefficients $C_i(\muEW)$ and
$C^\prime_i(\muEW)$. That is, even if a given $P_i$ is non-zero, the fate of its
contribution depends on the difference of these two coefficients. In particular,
in models with exact left-right symmetry this contribution vanishes as first
pointed out in \cite{Branco:1982wp}.

The numerical values of the $P_i(\muEW)$ are collected in the tables in
\refapp{app:epe-MF}.
 As seen in (\ref{eq:master2}),
the $P_i$ depend on the hadronic matrix elements
$\langle O_j (\muLow) \rangle_I$ and the RG evolution factors
$p_{ij}^{(I)}(\muEW, \muLow)$.  The numerical values of the hadronic matrix
elements rely on lattice QCD in the case of SM operators  and DQCD in the case
of BSM operators as summarized above. Consequently, the uncertainties of the $P_i$ are of the order of $5-7 \%$ resulting from SM matrix elements and at the level of $20 \%$ coming from BSM matrix elements.

Inspecting the results in the tables in \refapp{app:epe-MF} the following comments are in order.
\begin{itemize}
\item The large $P_i$ values for operators with flavour content
$(\bar sd)(\bar uu)$ and $(\bar sd)(\bar dd)$
 in Class~A can be traced back to the large values
  of the matrix elements $\langle Q_{7,8}\rangle_2$, the dominant electroweak
  penguin operators in the SM, and the enhancement of the
  $I=2$ contributions relative to $I=0$ ones by $\omega\approx22$.
\item The small $P_i$ values in Class~B are due to the fact that
  they are all proportional to $\langle O_{8g} \rangle_0$, which has recently been
  found to be much smaller than previously expected \cite{Constantinou:2017sgv,
    Buras:2018evv}. Moreover, as $\langle O_{8g}\rangle_2=0$, all contributions
  in this class are suppressed by the factor $1/\omega$ relative to
  contributions from other classes.
\item The large $P_i$ values in Classes C--D can be traced back to the large
  hadronic matrix elements of scalar and tensor operators calculated recently in
  \cite{Aebischer:2018rrz}.  Due to the smallness of $\langle O_{8g} \rangle_0$,
  the contribution of the chromo-magnetic dipole operator in Classes C--D is negligible.
\item While the $I=0$ matrix elements of the operators in Class E cannot
  be expressed in terms of SM ones, the $I=2$ matrix elements can, and the
  large $P_i$ values can be traced back to the large SM matrix elements $\langle Q_{7,8}\rangle_2$.
\end{itemize}

%
\subsection{Matching from SMEFT onto $\Delta S = 1$ EFT}\label{sec:matching}

The SMEFT is a convenient description of BSM scenarios that feature a large gap
between the NP and the electroweak scales, $\muNP \gg \muEW$. This implies that there
are only the known SM fields below $\muNP$ and it is assumed that the Higgs doublet
is in the linear representation. The SM dimension-four Lagrangian is supplemented
by a tower of local operators
\begin{align}
  \label{eq:Lag-SMEFT}
  \mathcal L_\text{SMEFT} &
  = \mathcal{L}_{\text{dim}-4} + \sum_{k} \Wc{k} \Op{k}\,,
\end{align}
that are invariant under the SM gauge group $G_\text{SM} = SU(3)_c \otimes SU(2)_L
\otimes U(1)_Y$ to describe physics below $\muNP$ around $\muEW$.

The SMEFT operators and accordingly their Wilson coefficients are defined in terms
of the gauge and fermion fields in the unbroken phase of the SM, see also
\refapp{app:SMEFT-operators} for notation and definitions. In contrast to the
$\Delta S=1$ EFT discussed above, there is no preferred weak basis for the (massless)
fermion fields in SMEFT and the would-be mass basis is not $SU(2)_L$ invariant.
Instead, in the following we use the freedom of $SU(3)$-flavour rotations to work in
a weak basis where the running down-type quark mass matrix is diagonal at the electroweak
scale (cf.~\cite{Aebischer:2017ugx}).

At the electroweak scale $\muEW$, the matching of SMEFT at the dimension-six level will only
generate a subset of the $\Delta S = 1$ operators introduced in \refapp{app:fierz},
because the SM gauge group $G_\text{SM}$ is more restrictive than $SU(3)_c \otimes
U(1)_Q$. Since flavour is not conserved by the RG mixing under the $SU(2)_L$-gauge
and Higgs-Yukawa interactions the SMEFT operators cannot be classified in terms
of flavour quantum numbers. Nevertheless, it is instructive to consider which
operators in SMEFT contribute to $\Delta S=1$ transitions when matched at
tree level onto~$\mathcal{H}_{\Delta S = 1}^{(5)}$ at the scale $\muEW$.

The matching of SMEFT onto the $\Delta S = 1$ EFT with five active quark
flavours yields relations between Wilson coefficients\footnote{We denote Wilson
coefficients of SMEFT by caligraphic $\Wc{k}$ and of low-energy EFTs by $C_k$.}
of both EFTs \cite{Aebischer:2015fzz, Jenkins:2017jig}.
Here we
focus on effects from three classes of operators:
\begin{itemize}
  \item four-quark operators,
  \item $\psi^2 H^2 D$ operators describing modified $Z^0$ or $W^\pm$ couplings, and
  \item chromo-magnetic dipole operators.
\end{itemize}
The matching conditions for these operators are collected in
\refapp{app:SMEFT-matching}.
We omit the effects from semi-leptonic operators
that have
been analysed in the context of leptoquark models \cite{Bobeth:2017ecx}
and are expected to be constrained more
strongly by semi-leptonic kaon decays rather than $\epe$. Furthermore we omit
effects of purely bosonic operators, which can induce a sizable contribution to
$\epe$ through RG effects as discussed in \cite{Bobeth:2015zqa}.

A non-trivial consequence of SMEFT is that none of the operators $O_{SLR}^{u_i}$,
$O_{SLL}^{d_i}$, $O_{TLL}^{d_i}$, or their chirality- and colour-flipped counterparts,
are generated in the low-energy EFT in the tree-level matching of SMEFT four-quark operators.
The reason is that these operators conserve only electric charge, but not hypercharge.
Only the operator $\widetilde{O}_{SLR}^{u}$ eventually contributes to $\epe$, namely
through the right-handed $W^\pm$ coupling discussed in
\refapp{sec:ZW}. This contribution is not subject to the hypercharge constraint,
as it only arises after electroweak symmetry breaking.
Below~$\muEW$ this leads to vanishing Wilson coefficients of $9 + 9'$ linearly
independent operators in the $\Delta S = 1$ EFT with $\Nf = 5$, reducing the
number of non-redundant $\Delta S=1$ four-quark operators that contribute to $\epe$
from $36 + 36'$ to $27 + 27'.$\footnote{%
Note that \refeq{eq:match1}--\refeq{eq:matchlast} contain operators that are
related by Fierz transformations. These allow to remove
12 of the 68 operators for $\Nf=5$ appearing on the left-hand side of these equations.}
At the one-loop level, QCD and QED running from $\muEW$ down to $\muLow$ does not
re-generate these operators. This is summarized in \reftab{tab:op-counting}.
Consequently, in SMEFT the number of linearly independent operators that contribute
directly to $\epe$ via non-vanishing $K\to\pi\pi$ matrix elements is reduced from
$15 + 15'$ to $12 + 12'$, out of which only $5 + 5'$  are
non-standard. The chromo-magnetic dipole operators are not subject to these
considerations and their number equals in SMEFT and $\Delta S = 1$ EFT.

\begin{table}
\centering
\renewcommand{\arraystretch}{1.3}
\begin{tabular}{lcccccc}
\toprule
          & $u_i = u$  & $u_i = c$ & $d_i = d$ & $d_i=s$  & $d_i = b$  & $\Sigma$      \\
\hline
  $\Delta S = 1$ EFT
          & $10 + 10'$ & $8 +  8'$ & $5 + 5'$  & $5 + 5'$ & $8 + 8'$ & $36 + 36'$ \\
  SMEFT   & $9  +  9'$ & $8 +  8'$ & $3 + 3'$  & $3 + 3'$ & $4 + 4'$ & $27 + 27'$ \\
\bottomrule
\end{tabular}
\caption{Number of linearly independent four-quark operators with flavour
  content $(\bar sd)(\bar u_i u_i)$ and $(\bar sd)(\bar d_i d_i)$ in the
  $\Delta S = 1$ EFT with $\Nf=5$ that contribute to $\epe$ (first row).
  Number of non-vanishing matching contributions from SMEFT due to four-quark
  operators and modified right-handed $W^\pm$ couplings for dimension-six
  operators at tree level (second row).}
  \label{tab:op-counting}
\end{table}

Consequently, in SMEFT only the operators in Classes A--C in \refeq{A1}--\refeq{C1}
contribute to $\epe$ through four-quark operators, and a single operator from
Class E in \refeq{E1} (and its chirality-flipped counterpart) through the right-handed
$W^\pm$ coupling. Inspecting the
matching relations listed in \refapp{app:SMEFT-matching}, these three classes, expressed in terms of the
SMEFT operators of \reftab{tab:SMEFT-4Fops} and \reftab{tab:SMEFT-magnops}, are as follows

\paragraph{Class A}

\begin{gather}
  \Op[(1,3)]{qq}\,, \qquad \Op[(1,8)]{qu}\,, \qquad
  \Op[(1,8)]{qd}\,, \qquad \Op[(1,8)]{ud}\,, \qquad \Op[]{dd}\,,
  \nonumber\\
  \Op[(1,3)]{Hq}\,, \qquad \Op[]{Hd}\,.
\end{gather}

\paragraph{Class B and C}

\begin{equation}\label{BC}
  \Op{dG}\,, \qquad \Op[(1,8)]{quqd}\,.
\end{equation}

\paragraph{Class E}

\begin{equation}\label{E11}
\Op[]{Hud}\,.
\end{equation}

It should be noted that while the matching conditions in \refsec{sec:matching}
are at the electroweak scale, the SMEFT operators are generated by some BSM
dynamics at a much higher scale $\muNP$ and in explicit models RG evolution in
the SMEFT from $\muNP$ to $\muEW$ has to be considered \cite{Jenkins:2013zja,
Jenkins:2013wua, Alonso:2013hga}. The RG evolution does not only change the
values of the Wilson coefficients through self-mixing of a given operator
but also through mixing of other operators, in particular those that do not
contribute directly to $\epe$ at tree-level. The mixing is further complicated
due to the flavour structure of the ADMs and can give rise to complex
correlation patterns between observables of various quark and lepton flavour
sectors. In appendix~\ref{app:SMEFT-running}, we list the classes of operators
that mix into operators contributing to the tree-level matching onto the
$\Delta S=1$ EFT and can thus be relevant for $\epe$.

%
%
%
\section{Model-independent constraints in SMEFT}\label{sec:constraints}

In specific NP models, CP-violating effects that can manifest themselves
in $\epe$ are constrained by other CP-odd observables, such as in
$\Delta S=2$ ($K^0$-$\bar K^0$ mixing),
$\Delta C=2$ ($D^0$-$\bar D^0$ mixing),
in electric dipole moments (EDM),
or in semi-leptonic kaon decays.
In the low-energy EFT, such effects cannot be discussed on a model-independent
basis, since the operators with different flavour quantum numbers are
completely independent. In SMEFT however, such correlations can
arise in two different ways,
\begin{itemize}
\item by $SU(2)_L$ relations between operators involving left-handed quark
doublets that require a CKM rotation
to go to the mass basis for the up- or down-type quarks,
\item by flavour-dependent RG effects due to the mixing of operators
  in SMEFT given in \refsec{sec:matching}.
\end{itemize}
In this section, we concentrate on effects of the first type, leading to
\textit{model-independent} constraints on Wilson coefficients of operators
contributing to $\epe$.

%
%
\subsection{$\Delta S=2$} \label{sec:DS2}

The parameter $\epsK$ measures \textit{indirect} CP violation in the $\Delta S=2$
process of $K^0$-$\bar K^0$ mixing. At the electroweak scale only
four linear combinations of the five\footnote{%
  We do not count redundant operators such as
  $\op[(1,8)]{qd}{1212} \equiv \op[(1,8)]{qd}{2121}^\dagger$, but adopt the
  basis of non-redundant operators defined in \cite{Celis:2017hod}.
  Contributions from $\psi^2 H^2 D$ operators corresponding to modified $Z^0$
  and $W^\pm$ couplings
  arise at one-loop from top-quark Yukawa mixing \cite{Bobeth:2017xry} and
  those from $h^0$ couplings count as beyond dimension six \cite{Jenkins:2017jig}.}
SMEFT operators
\begin{align}
  \label{eq:DS2-SMEFT-ops}
  \op[(1)]{qq}{2121} \,, \; \op[(3)]{qq}{2121} \,, & &
  \op[(1)]{qd}{2121} \,, \; \op[(8)]{qd}{2121} \,, & &
  \op[]{dd}{2121} \,,
\end{align}
match onto the $\Delta S=2$ EFT at tree level in the weak basis in which the
down-type quark mass matrix is diagonal.
The other four operators present in the $\Delta S=2$ EFT violate hypercharge
and are thus not generated at tree level \cite{Aebischer:2015fzz}.

The quantitative effect of these operators can be understood by writing $\epsK$ as a
function of their Wilson coefficients with approximate numerical coefficients,
\begin{align}
  \label{eq:epsnum}
  \frac{\epsK}{\epsK^\text{SM}} &
  \approx 1 + \sum_i \sigma_i \, \Lambda_i^2 \, \text{Im}\,\mathcal{C}_i(\muEW)\,,
\end{align}
where $\sigma_i=\pm1$.
Similarly to the $P_i$ in the master formula \refeq{eq:master} for $\epe$,
the effective scales $\Lambda_i$
give an indication of the sensitivity of $\epsK$ to each Wilson
coefficient; we list their numerical values in
\reftab{tab:eps}.
They have been obtained with
\texttt{flavio}~\cite{flavio} and \texttt{wilson}~\cite{Aebischer:2018bkb}
using the $\Delta S=2$ hadronic matrix elements
from lattice QCD by RBC-UKQCD \cite{Garron:2016mva, Boyle:2017skn}
(cf.\ results from the ETM \cite{Carrasco:2015pra}
and SWME \cite{Jang:2015sla} collaborations),
which are supported by DQCD results \cite{Buras:2018lgu}.

As the SM describes the experimental value of $\epsK$ rather
well, $\text{Im}\,\mathcal{C}_i(\muEW)$ corresponding to the largest  $\Lambda_i$
must be suppressed most strongly, thereby probing the largest NP scales.
Given the experimental measurement and theory uncertainty of this ratio,
\refeq{eq:epsnum} can be used to constrain
SMEFT Wilson coefficients from $\epsK$
in phenomenological analyses.

\begin{table}
\centering
\renewcommand{\arraystretch}{1.4}
\begin{tabular}{ccrcrrcrr}
\toprule
  $\mathcal{C}_i$ & $\sigma_i$ & $\Lambda_i$
& $\mathcal{C}_i$ & $\sigma_i$ & $\Lambda_i$
& $\mathcal{C}_i$ & $\sigma_i$ & $\Lambda_i$
\\
\midrule
  \multicolumn{9}{c}{$\Delta S = 2$}
\\
\midrule
  $\wc[(1)]{qq}{2121}$ & $-$ &  $13.3$ PeV
& $\wc[(1)]{qd}{2121}$ & $+$ &  $104.6$ PeV
& $\wc[]{dd}{2121}$    & $-$ &  $13.3$ PeV
\\
  $\wc[(3)]{qq}{2121}$ & $-$ &  $13.3$ PeV
& $\wc[(8)]{qd}{2121}$ & $+$ &  $126.5$ PeV
&
\\
\midrule
  \multicolumn{9}{c}{$\Delta C = 2$}
\\
\midrule
  $\wcHat[(1)]{qq}{1212}$ & $-$ & $14.1$ PeV
& $\wcHat[(1)]{qu}{1212}$ & $+$ & $29.2$ PeV
& $\wcHat[]{uu}{1212}$ & $-$ &  $14.1$ PeV
\\
  $\wcHat[(3)]{qq}{1212}$ & $-$ &  $14.1$ PeV
& $\wcHat[(8)]{qu}{1212}$ & $+$ &  $33.3$ PeV
&
\\
\bottomrule
\end{tabular}
\renewcommand{\arraystretch}{1.0}
  \caption{Effective scales of SMEFT operators contributing to $\epsK$ and
  CP violation in $D^0$-$\bar D^0$ mixing,
  defined as in~\refeq{eq:epsnum} and \eqref{eq:x12num}, respectively.
  These scales give an indication of the sensitivity to the individual operators.
  Note however that the normalization is different for $\Delta S=2$
  and $\Delta C=2$.}
  \label{tab:eps}
\end{table}

Given these huge scales probed by $\epsK$, any model predicting sizable
direct CP violation in $\Delta S=1$ can only be viable if it does not
induce too large contributions to indirect CP violation in $\Delta S=2$.

As discussed above, an important source of constraints
are $SU(2)_L$ relations between operators with
left-handed quark fields, involving a CKM rotation between the mass bases for
up- and down-type quarks.
The first four operators in \reftab{tab:eps} are a prime example of this effect.
$\wc[(1,3)]{qq}{2121}$ contribute to the matching of $C_{VLL}^{u_i}$ and
$\wc[(1,8)]{qd}{2121}$ to the matching of $C_{VRL}^{u_i}$,
as seen from the matching conditions in \refsec{sec:matching}.
For both cases $i = 1, 2$, the suppression by the Cabibbo angle $V_{us}^{}
V_{ud}^* \sim V_{cs}^{} V_{cd}^* \sim 0.23$ is of first order in
$\epe$ and furthermore the operators with $i = 2$ have no direct $K\to\pi\pi$
matrix elements, which introduces for them another suppression of
$\alpha_{s,e}/(4\pi)$ from RG mixing in $\epe$ compared to $i = 1$.
When considering NP effects in only a single operator, clearly the strong
constraint from $\epsK$ excludes any visible effect in $\epe$ induced by
imaginary parts of these Wilson coefficients.

We finally note that,
as emphasized in  \cite{Buras:2015jaq}, new phases could have an impact not only
on $\epsK$, but also on the mass difference in $K^0$-$\bar K^0$ mixing,
$\Delta M_K$. The point is that $\Delta M_K$ is proportional
to the real part of the square of a complex coefficient $C_i$, so a new phase modifying
its imaginary part will quite generally decrease the value of $\Delta M_K$ relative
to the SM estimate simply because
\begin{align}
  (\Delta M_K)_i^\text{BSM} &
  = c \left[(\text{Re}\,C_i)^2-(\text{Im}\,C_i)^2\right]\,,
\end{align}
with $c$ being positive. The uncertainty in the SM estimate of $\Delta M_K$ is
unfortunately still very large \cite{Brod:2011ty} so that we cannot presently
decide whether a positive or negative NP contribution to $\Delta M_K$ -- if any --
is required and the constraints on the NP scale are weaker than for $\epsK$. Future lattice QCD
calculations of long distance contributions to $\Delta M_K$ could help in this
respect \cite{Bai:2014cva, Christ:2015pwa}. In DQCD they are found to amount to
$20\pm 10\%$  of the measured $\Delta M_K$ \cite{Bijnens:1990mz, Buras:2014maa}.
In the case of  $\epsK$ such long distance contributions to $\epsK$
are below $10\%$ and have been reliably calculated in \cite{Buras:2008nn, Buras:2010pza,
Blum:2015ywa}.

%
%
\subsection{$\Delta C=2$}\label{sec:DC2}

Although the SM contribution to the $D^0$-$\bar D^0$ mixing amplitude is
dominated by poorly known long-distance contributions, the structure of the CKM
matrix implies that the SM contribution to CP violation in mixing can at most
reach the percent level \cite{Bobrowski:2010xg}. This fact can be used to
constrain the imaginary part of the mixing amplitude.

For processes with external up-type quarks, it is more convenient to use a
weak basis for SMEFT Wilson coefficients where the up-type rather than the
down-type quark mass matrix is diagonal.\footnote{This basis is denoted
\texttt{Warsaw up} in the WCxf standard \cite{Aebischer:2017ugx}.} We will
denote the Wilson coefficients in this basis with a hat.
The hatted Wilson coefficients are related to the unhatted ones
by CKM rotations of indices corresponding to left-handed quark doublets,
\begin{equation}
  \label{eq:def-hatted-wc}
\begin{aligned}
  \wcHat[(1,3)]{qq}{ijkl} &=
  V_{ia}^{} \, V_{jb}^* \, V_{kc}^{} \, V_{ld}^* \,  \wc[(1,3)]{qq}{abcd}
  \,,\\
  \wcHat[(1,8)]{qu}{ijkl} &=
  V_{ia}^{} \, V_{jb}^* \,  \wc[(1,8)]{qu}{abkl}
  \,,\\
  \wcHat[]{uu}{ijkl} &=
  \wc[]{uu}{ijkl}
  \,.
\end{aligned}
\end{equation}
Then, analogously to $\epsK$, at the electroweak scale
four linear combinations of
five SMEFT operators contribute to $\Delta C=2$ transitions, namely
\begin{align}
  \opHat[(1)]{qq}{1212} \,, \; \opHat[(3)]{qq}{1212} \,, & &
  \opHat[(1)]{qu}{1212} \,, \; \opHat[(8)]{qu}{1212} \,, & &
  \opHat[]{uu}{1212} \,.
  \label{eq:df2ops}
\end{align}
A correlation of $\epe$ and $D^0$-$\bar D^0$ mixing arises only for the operators
$\Op[(1,3)]{qq}$ and $\Op[(1,8)]{qu}$ as can be seen from
\refeq{eq:match1}--\refeq{eq:matchlast}.

From a global fit to $D^0$ decays, the HFLAV collaboration directly determines
the physical parameters of the $D^0$-$\bar D^0$ mixing amplitude,
\begin{align}
  x_{12} &= \frac{2|M_{12}|}{\Gamma}
  \,,&
  y_{12} &= \frac{|\Gamma_{12}|}{\Gamma}
  \,,&
  \phi_{12} &= \text{arg}\frac{M_{12}}{\Gamma_{12}}
  \,.
\end{align}
Their fit result can be expressed as an approximately Gaussian constraint on
the purely CP-violating parameter \cite{Amhis:2016xyh}
\begin{equation}
x_{12}^\text{Im} \equiv x_{12} \sin \phi_{12} = (0 \pm 2.4) \times 10^{-4}.
\end{equation}
Similarly to the discussion of $\epsK$ above, we can write $x_{12}^\text{Im}$
as a linear function of SMEFT Wilson coefficients at $\muEW$,
\begin{equation}
\frac{x_{12}^\text{Im}}{10^{-4}} \approx
\sum_i \sigma_i \Lambda_i^2 \, \text{Im}\,\widehat{\mathcal{C}}_i \,.
\label{eq:x12num}
\end{equation}
The effective sensitivity scales $\Lambda_i$ are given in \reftab{tab:eps}.
They have been evaluated with
\texttt{flavio}~\cite{flavio} and \texttt{wilson}~\cite{Aebischer:2018bkb}
using the $\Delta C=2$ hadronic matrix elements
from lattice QCD by the ETM collaboration \cite{Carrasco:2015pra}.

Similarly to the $\Delta S=2$ case, we see that the four Wilson coefficients
$\wcHat[(1,3)]{qq}{1212}$ and $\wcHat[(1,8)]{qu}{1212}$
individually cannot give a visible effect in $\epe$ without
generating excessive contributions to CP violation in $\Delta C=2$.

%
%
\subsection{Interplay of $\Delta S=2$ and $\Delta C=2$}\label{sec:DF2}

\begin{figure}[tbp]
\centering
\includegraphics[width=\textwidth]{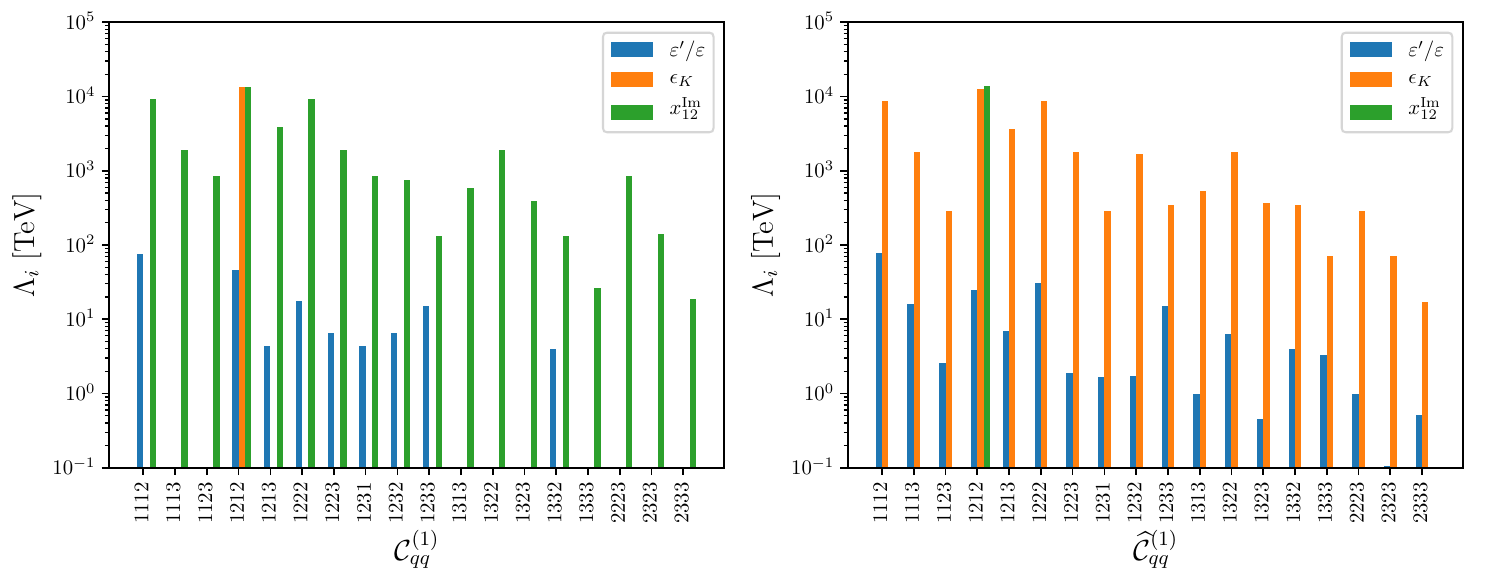}
\caption{Effective scales $\Lambda_i$ of the Wilson coefficients $\Wc[(1)]{qq}$
  for $\epsK$ (orange), $D^0$-$\bar D^0$ mixing (green) and the NP contribution
  to $\epe$ (blue), parametrized as in \eqref{eq:epsnum}, \eqref{eq:x12num},
  and~\eqref{eq:epenum-qq}, respectively.
  The left panel shows the values in the ``unhatted'' basis where the down-type
  quark mass matrix is diagonal, the right panel in the ``hatted'' basis where the
  up-type quark mass matrix is diagonal. Only non-redundant flavour-index combinations
  are shown. Coefficients that do not generate a visible effect in either observable
  have been omitted. The scales corresponding to $\Wc[(3)]{qq}$ are
  not shown but are very similar.}
\label{fig:df2}
\end{figure}

While individually, $\Delta S=2$ and $\Delta C=2$ only constrain seven
linear combinations of SMEFT Wilson coefficients contributing
to the $\Delta S=1$ matching, combining them leads to a much more powerful
constraint. This is because the Wilson coefficients
$\wc[(1,3)]{qq}{1212}$ and $\wcHat[(1,3)]{qq}{1212}$ are
related by CKM rotations:
\begin{align}
  \wcHat[(1,3)]{qq}{1212} & = V_{ui}^{}\,V_{cj}^* \,V_{uk}^{}\,V_{cl}^*  \,\wc[(1,3)]{qq}{ijkl}
  \,,\\
  \wc[(1,3)]{qq}{1212}    & = V_{id}^* \,V_{js}^{}\,V_{kd}^* \,V_{ls}^{} \,\wcHat[(1,3)]{qq}{ijkl}
  \,.
\end{align}
Consequently, for any given operator of this type, it is impossible to avoid both the contribution to $\Delta S=2$ and $\Delta C=2$ at the same time (cf.\ the general discussion in \cite{Blum:2009sk}).
Indeed, switching on individual operators in either of the two bases at $\muEW$, it turns out they all lead to an excessive contribution to either $\epsK$ or $x_{12}^\text{Im}$ when generating a visible effect in $\epe$.
This is illustrated in \reffig{fig:df2},
showing the suppression scales $\Lambda_i$ for $\epsK$ and  $x_{12}^\text{Im}$ (as defined in \eqref{eq:epsnum},
\eqref{eq:x12num}) and comparing it to the analogous scale for $\epe$,
defined as
\begin{equation}
\frac{(\epe)_\text{BSM}}{10^{-3}} \approx
\sum_i \sigma_i \Lambda_i^2 \, \text{Im}\,\wc[(1,3)]{qq}{i} \,,
\label{eq:epenum-qq}
\end{equation}
in the two different
bases where either the down-type or the up-type quark mass matrix is diagonal. While in the former basis $\epsK$ and in the latter basis
$x_{12}^\text{Im}$ is only sensitive to a single coefficient,
the other observable probes all the other coefficients,
always being much more sensitive than $\epe$.

We finally note that, in principle, since each of the observables only probes a
single direction in the space of Wilson coefficients, cancellations could be
arranged that remove these constraints. In view of the severeness of the constraints
and the fact that delicate cancellations are not invariant under the RG evolution,
we consider such cancellations unrealistic.

%
%
\subsection{Neutron electric dipole moment}\label{sec:edm}

Since $\epe$ probes CP violation associated to the first two
generations of quarks, it is natural to ask whether there is
any constraint from the electric dipole moment (EDM) of the neutron,
which is a sensitive probe of \textit{flavour-diagonal} CP violation
involving up and down quarks. In principle, CP-violating four-quark
operators can directly induce a neutron EDM.
Correlations of
the neutron EDM with $\epe$ from these operators have been considered
recently in \cite{Cirigliano:2016yhc, Haba:2018byj};
they require the knowledge of the matrix elements of these
operators, which are relatively poorly known.

Here we focus instead on CP violation
induced by dipole operators, i.e.\ the EDMs and chromo-EDMs (CEDMs) of the up and
down quarks. Their contribution to the neutron EDM can be written
as\footnote{We neglect a numerically subleading part from the strange quark,
since $g_T^s \ll g_T^{u,d}$, and assume that the contribution from
the strange quark CEDM can be neglected as well.}
\begin{align}
  d_n  &
  = g_T^u \, d_u + g_T^d \, d_d
  + \tilde{\rho}_u \tilde{d}_u + \tilde{\rho}_d \tilde{d}_d \,.
\end{align}
The tensor charges $g_T^{u,d}$ are nowadays accessible in lattice QCD with an accuracy
of (5--10)\% \cite{Bhattacharya:2015wna, Alexandrou:2017qyt, Yamanaka:2018uud, Gupta:2018lvp},
while the matrix elements $\tilde{\rho}_{u,d}$ of the CEDMs are only known roughly
from methods like light-cone sum rules \cite{Pospelov:2000bw, Fuyuto:2013gla}.
The quark (C)EDMs are simply the imaginary parts of the Wilson coefficients
of the flavour-diagonal dipole operators at the hadronic scale\footnote{
Note that the signs on the right-hand sides of \refeq{eq:edm} depend on the
sign convention for the covariant derivative. We use
$D_\mu=\partial_\mu + ieQ_fA_\mu +ig_s G_\mu^a T^a$. Our convention for $\sigma^{\mu\nu}$
is $\sigma^{\mu\nu}=\frac{i}{2}[\gamma_\mu,\gamma_\nu]$.
},
\begin{align}
  d_q &= 2 m_q ~\text{Im}\, C_{7\gamma}^{qq}
\,,&
g_s \, \tilde d_q &= 2 m_q ~\text{Im}\, C_{8g}^{qq}
\,,
\label{eq:edm}
\end{align}
with the effective Hamiltonian
\begin{align}
\mathcal H_{\Delta F=0} &
 = - \sum_{q=u,d} \left[
   C_{7\gamma}^{qq} \, m_q(\bar{q} \sigma^{\mu\nu} P_R q) F_{\mu\nu}
 + C_{8g}^{qq}      \, m_q(\bar{q} \sigma^{\mu\nu} P_R T^A q) G^A_{\mu\nu}
 + \text{h.c.}
\right].
\end{align}
Below the electroweak scale, the dipole operators receive RG-induced
contributions via QCD and QED penguin diagrams from operators with
chirality structure LRLR,
\begin{align}
  O_{XAA}^{qqpp} &
  = (\bar q^i \Gamma_X P_A q^i) (\bar p^j \Gamma_X P_A p^j)
\,,&
  \widetilde{O}_{XAA}^{qqpp} &
  = (\bar q^i \Gamma_X P_A q^j) (\bar p^j \Gamma_X P_A p^i) \,,
\label{eq:opdf0}
\end{align}
where $X=S,T$ and $A=L,R$.
In tree-level matching from SMEFT at $\muEW$, such operators are only
generated from the SMEFT operators $\Op[(1,8)]{quqd}$, similarly to the
operators $C_{SAA}^{u_i}$ and  $C_{TAA}^{u_i}$ in the $\Delta S=1$
matching in section~\ref{sec:matching}. Via CKM rotations, many of the
operators in \eqref{eq:opdf0} are thus related to $\Delta S=1$ operators.

\begin{figure}[tbp]
\centering
\includegraphics[width=\textwidth]{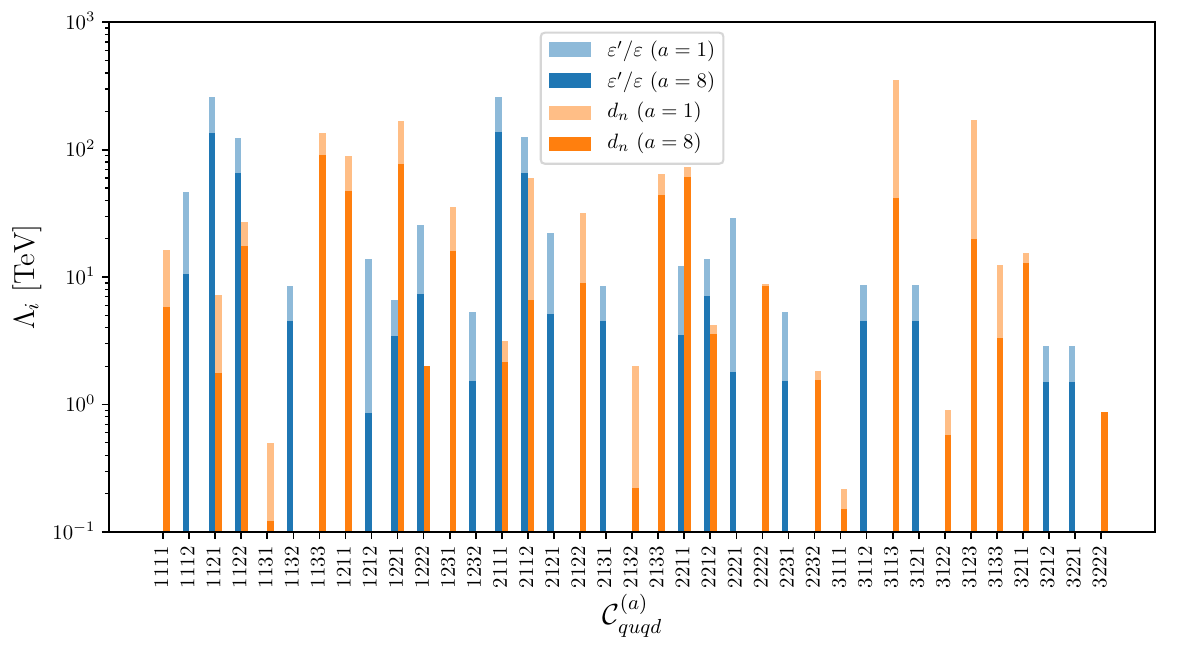}
\caption{Effective scales $\Lambda_i$ for the neutron EDM [orange] and the NP
contribution to $\epe$~[blue], parametrized as in \eqref{eq:dnnum} and
\eqref{eq:epenum}, respectively. Only non-redundant index combinations are
shown. Coefficients that do not generate a visible effect in either observable
have been omitted. The scales corresponding to $\Wc[(1)]{quqd}$ are shown
with a lighter shading than $\Wc[(8)]{quqd}$ and are always higher.}
\label{fig:edm}
\end{figure}

Analogously to the discussion of $\epsK$ and $D^0$-$\bar D^0$ mixing,
the constraints on the operators $\Op[(1,8)]{quqd}$ from $d_n$ can be illustrated
by writing $d_n$ as a linear combination of Wilson coefficients at $\muEW$,
\begin{equation}
\frac{d_n}{d_n^\text{lim}} \approx
\sum_i \sigma_i \Lambda_i^2 \, \text{Im}\,\wc[(1,8)]{quqd}{i} \,,
\label{eq:dnnum}
\end{equation}
where $i$ stands for a 4-tuple of flavour indices and
$d_n^\text{lim}$ is the current 90\% confidence-level upper bound
on the neutron EDM \cite{Afach:2015sja},
\begin{equation}
d_n^\text{lim} = 3 \times 10^{-26} ~e\,\text{cm} \approx 4.6 \times 10^{-13} ~\text{GeV}^{-1} \,.
\end{equation}
In \reffig{fig:edm}, we show the values of $\Lambda_i$ for the neutron EDM
(obtained with
\texttt{flavio}~\cite{flavio} and \texttt{wilson}~\cite{Aebischer:2018bkb})
as well as for $\epe$, parametrized analogously as
\begin{equation}
\frac{(\epe)_\text{BSM}}{10^{-3}} \approx
\sum_i \sigma_i \Lambda_i^2 \, \text{Im}\,\wc[(1,8)]{quqd}{i} \,.
\label{eq:epenum}
\end{equation}
The chart shows that several of the operators would lead to an excessive
contribution to $d_n$ when leading to a visible effect in $\epe$;
some of them do not contribute to $\epe$; and yet others can generate $\epe$
without being constrained by $d_n$. We have omitted the operators that do not
contribute to either of the observables.

We stress again that the correlation discussed here arises simply from CKM
rotations when moving between the mass bases of up and down quarks
and we have considered SMEFT Wilson coefficients at $\muEW$.
When considering the coefficients at a high scale $\muNP$,
there are also RG effects in SMEFT that induce mixing between $\Op[(1,8)]{quqd}$
with different flavour indices that can lead
to additional dangerous contributions to $d_n$.
Whether a visible NP effect in $\epe$ generated by any of the operators
$\Op[(1,8)]{quqd}$ is viable in view of the EDM constraint has to be checked
carefully in specific NP models taking into account both effects.

We finally note that beyond the neutron EDM, also the EDMs of diamagnetic atoms are
sensitive to CP violation in dipole operators and four-quark operators,
in addition to leptonic and semi-leptonic CP violation.
In principle a global analysis of the various EDM measurements
to disentangle the different short-distance sources of CP violation
would be useful, but currently suffers from many unknown long-distance contributions,
see \cite{Yamanaka:2017mef} for a recent review.

%
%
\subsection{$K\to\pi\nu\bar\nu$ and $K\to\pi\ell^+\ell^-$}\label{sec:sl}

In specific NP models one often finds correlations between BSM contributions
to $\epe$ and rare kaon decays, in particular with $\kpn$ and $\klpn$. In
fact in all papers that addressed the $\epe$ anomaly listed in the introduction
\cite{Buras:2015kwd, Buras:2016dxz, Bobeth:2017xry, Endo:2016tnu, Bobeth:2016llm,
Blanke:2015wba, Buras:2015yca, Buras:2015jaq, Tanimoto:2016yfy, Kitahara:2016otd,
Endo:2016aws, Crivellin:2017gks, Endo:2017ums, Chen:2018ytc, Chen:2018vog,
Haba:2018byj, Haba:2018rzf, Matsuzaki:2018jui, Bobeth:2017ecx} such correlations
have been investigated. Such correlations will play an important role in
distinguishing various models when the theoretical status of $\epe$ improves
and the branching ratios for rare kaon decays will be well measured.

Here
we would like to confine our discussion to possible model-independent
correlations within a pure EFT analysis.
Correlations between $\epe$ and semi-leptonic decays can then
in principle arise in three different ways,
\begin{itemize}
  \item modified $Z^0$ or $W^\pm$ couplings contributing to $\epe$ and
  neutral or charged current semi-leptonic decays, respectively,
  \item semi-leptonic operators that contribute directly to semi-leptonic decays
  and mix into $\Delta S=1$ four-quark operators by QED or electroweak RG effects,
  thereby contributing indirectly to $\epe$,
  \item four-quark operators mixing into semi-leptonic operators
  by QED or electroweak RG effects and contributing directly to $\epe$.
\end{itemize}
The latter two effects are strongly suppressed by the smallness of the
electroweak gauge couplings; consequently $\epe$ typically dominates
constraints on CP violation in four-quark operators, while semi-leptonic decays dominate
constraints on semi-leptonic operators.

Relevant model-independent correlations could thus arise from
the modified $Z^0$ or $W^\pm$ couplings induced by the SMEFT operators
of type $\psi^2H^2D$ discussed in \refsec{sec:ZW}. From the discussion in that
section, it was concluded that imaginary parts of the following SMEFT Wilson
coefficients at $\muEW$ can lead to effects in $\epe$,
\begin{gather}
  \wc[]{Hd}{12}
  \,,\qquad
  \wc[(1)]{Hq}{12}
  \,,\qquad
  \wc[(3)]{Hq}{12}
  \,,\label{eq:CHx1} \\
  \wc[(3)]{Hq}{13}
  \,,\qquad
  \wc[(3)]{Hq}{23}
  \,,\label{eq:CHx2} \\
  \wc{Hud}{12}
  \,,\qquad
  \wc{Hud}{11}
  \,.\label{eq:CHx3}
\end{gather}
The coefficients of right-handed $W^\pm$ couplings in
\refeq{eq:CHx3} contribute at tree-level only to {\em charged-current}
semi-leptonic decays like
$K\to \ell\nu_\ell$,
$K\to\pi\ell\nu_\ell$,
and beta decays
(see e.g.~\cite{Cirigliano:2013xha,Gonzalez-Alonso:2016etj,Gonzalez-Alonso:2018omy})
and the effects in $\epe$ are essentially
unconstrained at present.

The coefficients in \refeq{eq:CHx2}, which contribute to $\epe$
only via modified left-handed $W^\pm$ couplings,
contribute also to FCNC $B$ decays via modified $Z^0$ couplings.
Barring unrealistic cancellations,
visible effects in $\epe$ induced by these couplings
are excluded since they would lead to excessive
effects e.g.\ in the decays $B_s\to\mu^+\mu^-$ and  $B^0\to\mu^+\mu^-$.

\begin{figure}[tbp]
\centering
\includegraphics[width=\textwidth]{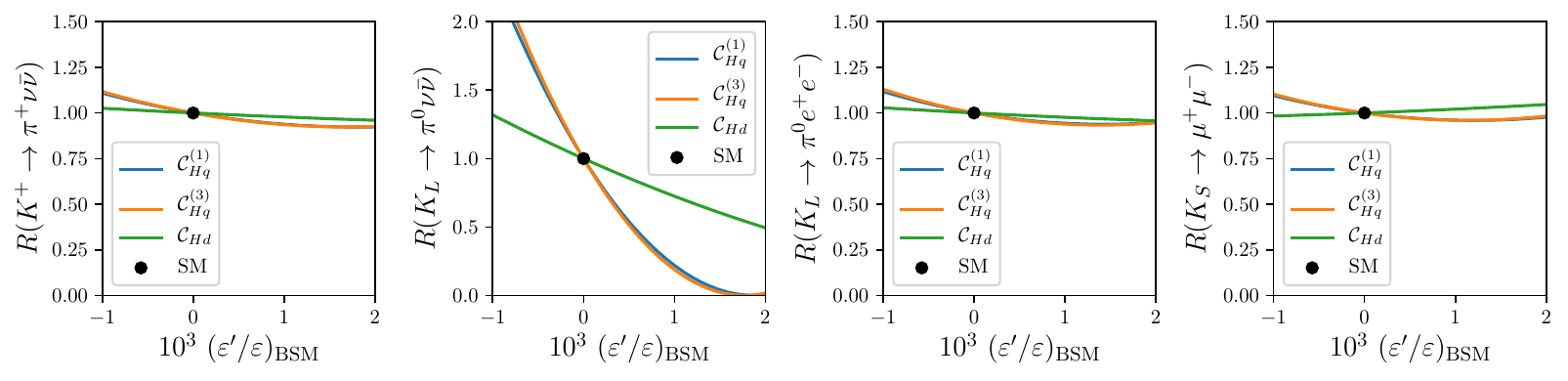}
\caption{Correlation between $\epe$, $\kpn$, $\klpn$, $\klpee$, and $K_S\to\mu^+\mu^-$
from imaginary NP effects in individual SMEFT operators of
type $\psi^2 H^2 D$ inducing flavour-changing $Z^0$ couplings.
The flavour index ``12'' has been suppressed.}
\label{fig:sl}
\end{figure}

The coefficients in \refeq{eq:CHx1} contribute to the FCNC kaon decays
of type $K\to\pi\nu\bar\nu$ and $K\to\pi\ell^+\ell^-$. These decays are
sensitive to a single linear combination, namely
\begin{equation}
   \wc[(3)]{Hq}{12}  + \wc[(1)]{Hq}{12}  + \wc{Hd}{12}  \,,
\end{equation}
while the leptonic FCNC decays of type $K\to\ell^+\ell^-$
are sensitive to
\begin{equation}
   \wc[(3)]{Hq}{12}  + \wc[(1)]{Hq}{12}  - \wc{Hd}{12}  \,.
\end{equation}
Inspecting our master formula and matching conditions, $\epe$ is
instead sensitive approximately to the imaginary
part of the linear combination
\begin{equation}
  \wc[(3)]{Hq}{12} + 1.1\,\wc[(1)]{Hq}{12} +3.7\, \wc[]{Hd}{12}
\end{equation}
of these three Wilson coefficients at $\muEW$.
Numerically, it turns out that
a purely CP violating contribution to any of these three coefficients
that would lead to a visible effect in $\epe$ only leads to a very small
modification of the $\kpn$, $\klpll$, and
$K_S\to\ell^+\ell^-$ branching ratios, as
demonstrated in \reffig{fig:sl} (see also \cite{Bobeth:2016llm, Bobeth:2017xry}).
In the CP violating decay $\klpn$, a NP effect in $\epe$ in the ballpark of
$10^{-3}$ would instead lead to a suppressed branching ratio. Seeing such
suppression would however require an experimental sensitivity better than the
SM branching ratio, which is at the
level of $3\times 10^{-11}$, still two orders of magnitude away
from the recent preliminary bound from the KOTO
collaboration\cite{koto-ichep18},
\begin{equation}
  \text{BR}(\klpn) < 3.0 \times 10^{-9} \qquad\text{@ 90\% C.L.}
\end{equation}

We conclude that CP-violating new physics in the operators with modified
$Z^0$ couplings in~\eqref{eq:CHx1}
or right-handed $W^\pm$ couplings in~\eqref{eq:CHx3}
can lead to
sizable effects in $\epe$ without appreciable constraints from semi-leptonic
kaon decays.

%
%
\subsection{$\Delta C = 1$}

Eventually we mention that similarly to $\Delta C = 2$ processes,
also CP violation in $\Delta C = 1$ decays is correlated to $\epe$ in SMEFT.
The interesting observables are CP asymmetries in Cabbibo-favoured~(CF) and
singly-Cabbibo suppressed~(SCS) $D \to M_1 M_2$ decays. They are governed
by the $\Delta C = 1$ EFT
\begin{align}
  \mathcal{H}_{\Delta C = 1} &
  = - \sum_i C_i O_i\,,
\end{align}
with operators
\begin{align}
   O_{XAB}^{qq'} &
   = (\bar u^i \Gamma_X P_A c^i) (\bar q^j \Gamma_X P_B q^{\prime j}) \,,
& 
   \widetilde{O}_{XAB}^{qq'} &
   = (\bar u^i \Gamma_X P_A c^j) (\bar q^j \Gamma_X P_B q^{\prime i}) \,,
\end{align}
with $qq' = sd$ (CF) and $qq' = dd, ss$ or $qq' = uu$ (SCS). The correlations
enter then via the SMEFT four-quark operators
\begin{align}
  \label{eq:DC1-SMEFT-ops}
  \Op[(1)]{qq} \,, \; \Op[(3)]{qq} \,, & &
  \Op[(1)]{qu} \,, \; \Op[(8)]{qu} \,, & &
  \Op[(1)]{qd} \,, \; \Op[(8)]{qd} \,, & &
  \Op[(1)]{quqd} \,, \; \Op[(8)]{quqd} \,,
\end{align}
as well as modified $Z^0$ and $W^\pm$ couplings.

The correlation
of $\epe$ and CP asymmetries in CF decays $D^0\to K^-\pi^+$, $D_s^+\to \eta \pi^+$
and
$D_s^+\to \eta' \pi^+$ has been discussed \cite{Haba:2018rzf} in the framework
of a $SU(2)_L \otimes SU(2)_R \otimes U(1)_{B-L}$ symmetric model.
The correlation with CP asymmetries in the SCS decays $D^0\to K^+K^-$
and $D^0\to\pi^+\pi^-$ has been discussed in a general EFT framework in
\cite{Isidori:2011qw}.
The rich potential to distinguish among various BSM scenarios with the
help of these correlations is hampered by the lack of knowledge of hadronic
matrix elements in non-leptonic charm decays and we will therefore not investigate
this subject further.

%
%
%
\section{Implications for model building}\label{sec:Lessons}

Having discussed the general model-independent anatomy of $\epe$ below the
electroweak scale and the consequences of $SU(2)_L\otimes U(1)_Y$ gauge
invariance within SMEFT, we are now in a position to discuss the implications
for the possible effects in BSM scenarios with new sources of CP violation where
BSM effects in $\epe$ are encoded in the imaginary part of Wilson coefficients
of dimension-six SMEFT operators.

The size of the coefficients $P_i$ in our master formula presented in
\refsec{sec:master}, together with the matching conditions in \refsec{sec:matching},
already indicate which scenarios are more promising than others to explain a
deviation from the SM in $\epe$.
However, in a concrete BSM scenario, the Wilson coefficients with the highest
values of $P_i$ could vanish or be suppressed by small couplings. Consequently
without additional dynamical assumptions or specific models no clear-cut
conclusions can be made.
While a comprehensive discussion of models is beyond the scope of this paper,
in the following subsections
we will discuss a number of general implications on the basis
of simplified models with a single tree-level mediator.

A generic challenge in explaining sizable NP effects in $\epe$ is to
avoid the constraint from $\epsK$. Roughly speaking, the $\Delta S=1$
CP-odd observable $\epe$ typically probes the quantity $\text{Im}\,\delta/\muNP^2$,
where $\delta$ is a flavour-violating parameter, while the $\Delta S=2$
observable $\epsK$ typically probes $\text{Im}\,\delta^2/\muNP^2$
(cf.~\cite{Kitahara:2016otd}). Given the strong constraints
from $\epsK$ (see \reftab{tab:eps}),
barring a tuning of the phase of $\delta$ or fine-tuned cancellations,
a visible effect in $\epe$ then seemingly requires
very low NP scales $\muNP \lesssim 1\,\text{TeV}$.
In the literature, this problem has been avoided in four different ways,
\begin{itemize}
\item through contributions from chromo-magnetic dipole operators to
  $\epe$ that do not affect~$\epsK$ \cite{Chen:2018vog},
\item through contributions from modified $Z^0$ couplings to
  $\epe$ \cite{Endo:2016tnu,Bobeth:2016llm,Blanke:2015wba} that only enter $\epsK$
  through top-quark Yukawa RG effects \cite{Bobeth:2017xry},
\item through contributions from modified right-handed $W^\pm$ couplings to
  $\epe$ that do not affect $\epsK$ \cite{Cirigliano:2016yhc},
\item through loop-induced contributions to $\epe$ in conjunction with an
accidental suppression of the contributions to $\epsK$ arising in models with
Majorana fermions like the MSSM \cite{Kitahara:2016otd}.
\end{itemize}
In \refsec{sec:scalar}, we will present a new solution:
tree-level scalar exchange can mediate $\Delta S=1$ transitions at tree level
without generating $\Delta S=2$, since a dimension six operator of the
form $(\bar q d)^2$ is not allowed by hypercharge invariance.

We start by listing all the possible tree-level models in \refsec{sec:tree}.
After discussing the scalar scenario in  \refsec{sec:scalar}, we will comment
on the challenges of models with vector mediators in \refsec{sec:vector} and
discuss the generation of modified $Z^0$ and $W^\pm$ couplings in
\refsec{sec:WZmodels}.

%
%
\subsection{Tree-level mediators}\label{sec:tree}

The simplest models giving rise to a NP contribution to $\epe$ are models with
a single tree-level mediator generating a four-quark operator. Given the large
scales probed by $\epe$, clearly also models \textit{without} tree-level
FCNCs can give a sizable contribution to $\epe$. Nevertheless, the tree-level
models can serve as benchmark cases exhibiting generic features of larger classes
of models.

In \reftab{tab:models}, we list all the possible tree-level mediators that
can generate any of the four-quark operators that give a matching
contribution to $\Delta S=1$ at $\muEW$
\cite{deBlas:2017xtg}. We have omitted
states that permit baryon number violating couplings.\footnote{Even though not all
of them mediate proton decay at tree level, see e.g.\ \cite{Arnold:2012sd, Assad:2017iib}.}
These states are either $SU(3)_c$ triplets (leptoquarks) or sextets (diquarks),
and the former are popular scenarios to explain current anomalies in semi-leptonic
$B$ decays. Further we omitted the possibility of a heavy vector doublet
$(1, 2)_{\frac 12}$.
For a scalar mediator, the SM gauge quantum numbers then only allow two possible
representations: a heavy Higgs-like doublet under $SU(2)_L$ that is either a
singlet or an octet under $SU(3)_c$. For a vector mediator, there are six
possibilities, $SU(3)_c$ singlets or octets that are $SU(2)_L$ singlets or triplets.

\begin{table}[tbp]
  \centering
  \begin{tabular}{clccccccccccc}
  \toprule
    Spin & Rep. &
    $\mathcal{O}^{(1)}_{qq}$ &
    $\mathcal{O}^{(3)}_{qq}$ &
    $\mathcal{O}^{(1)}_{qu}$ &
    $\mathcal{O}^{(8)}_{qu}$ &
    $\mathcal{O}^{(1)}_{qd}$ &
    $\mathcal{O}^{(8)}_{qd}$ &
    $\mathcal{O}^{(1)}_{ud}$ &
    $\mathcal{O}^{(8)}_{ud}$ &
    $\mathcal{O}_{dd}$ &
    $\mathcal{O}^{(1)}_{quqd}$ &
    $\mathcal{O}^{(8)}_{quqd}$
    \\
    \midrule
    \multirow{2}{*}{0} & $\left(1,2\right)_{\frac 12}$ &
    &&$\times$&$\times$&$\times$&$\times$&&&&$\times$\\
    & $\left(8,2\right)_{\frac 12}$ &
    &&$\times$&$\times$&$\times$&$\times$&&&&&$\times$\\
    \midrule
    \multirow{6}{*}{1} & $\left(1,1\right)_0$ &
    $\times$&&$\times$&&$\times$&&$\times$&&$\times$\\
    & $\left(1,1\right)_1$ &
    &&&&&&$\times$&$\times$\\
    & $\left(8,1\right)_0$ &
    $\times$&$\times$&&$\times$&&$\times$&&$\times$&$\times$\\
    & $\left(8,1\right)_1$ &
    &&&&&&$\times$&$\times$\\
    \cdashlinelr{2-13}
    & $\left(1,3\right)_0$ &
    &$\times$\\
    & $\left(8,3\right)_{0}$ &
    $\times$ & $\times$ \\
    \bottomrule
  \end{tabular}
  \caption{Four-quark SMEFT operators containing down-type quarks
  generated by the exchange of scalar or vector mediators
  at tree level. The second column gives the representation under
  $G_\text{SM} = SU(3)_c \otimes SU(2)_L \otimes U(1)_Y$.
  }
  \label{tab:models}
\end{table}

Further tree-level contributions to $\epe$ can arise from models inducing
modified $W^\pm$ or $Z^0$ couplings and will be discussed in
\refsec{sec:WZmodels}.

%
%
\subsection{Scalar operators from scalar mediators}\label{sec:scalar}

The novel feature after the calculation of hadronic matrix elements of BSM
operators in \cite{Aebischer:2018rrz} is the importance of scalar and
tensor four-quark operators.
As indicated in \reffig{fig:sketch} and shown in \refsec{sec:matching},
these matrix elements are relevant in scenarios that generate the SMEFT
operators $\Op[(1,8)]{quqd}$ at the electroweak scale.
\refTab{tab:models} shows that these operators can be mediated at tree level
by heavy Higgs doublets, either a colour-singlet or a colour-octet Higgs.

Focusing on the colour-octet case
(and thereby avoiding discussions of a modified SM Higgs potential),
the Lagrangian necessary to generate the $\Op[(1,8)]{quqd}$ operators
can be written as
\begin{align}
  \mathcal L &=
-X_d^{ij} \,\bar q_i \,T^A d_j \,\Phi^A
-
X_u^{ij} \,\bar q_i \,T^A u_j  \,\tilde \Phi^A
+ \text{h.c.}
\label{eq:LagPhi}
\end{align}
Integrating out the heavy scalar leads to the following tree-level
matching conditions for the four-quark SMEFT operators at the matching scale
$\muNP$ \cite{deBlas:2017xtg}
\begin{align}
  \wc[(1)]{qu(qd)}{ijkl} &
  = -\frac{4}{3} \wc[(8)]{qu(qd)}{ijkl}
  = -\frac{2}{9} \, \frac{X_{u(d)}^{jk*} X_{u(d)}^{il}}{M_\phi^{2}}  \,,
&
 \wc[(8)]{quqd}{ijkl} &
 = \frac{X_u^{ij} X_d^{kl}}{M_\phi^{2}} \,.
 \label{eq:scalarmatch}
\end{align}

Importantly, to generate  $\Wc[(8)]{quqd}$, the presence of both Yukawa-like
couplings $X_u$ and $X_d$ is necessary. The model can thus contribute
to $\epe$ both through the left-right vector operators $\Op[(1,8)]{qu,qd}$ and
through the scalar operators; which one is more relevant depends on
the hierarchies of the CP-violating couplings.

Some of the operators in \refeq{eq:scalarmatch} are also constrained by
the $\Delta F=2$ or $\Delta F=0$ processes discussed in
\refsec{sec:constraints}. In the basis where the down-type quark
mass matrix is diagonal, $\epsK$ is sensitive to $\wc[(1)]{qd}{2121}$.
As seen from \refeq{eq:scalarmatch}, this Wilson coefficient is proportional
to $X_d^{12*} X_d^{21}$. Interestingly, this means that an imaginary part
in one of the couplings $X_d^{12}$ or $X_d^{21}$ is not constrained by $\epsK$
at all, but could well generate a visible effect in $\epe$. Similar comments
apply to the $\Delta C=2$ constraint on flavour off-diagonal couplings in $X_u$.

Since the operators of type $\Op[(1,8)]{quqd}$ can be generated, in
models with scalar mediators also the neutron EDM, induced at
the one-loop level as discussed in \refsec{sec:edm}, can be a relevant
constraint.

We leave a detailed analysis of the interesting scalar scenarios to the
future.

%
%
\subsection{Models with vector mediators}\label{sec:vector}

As shown in \refsec{sec:DF2}, the operators
$\Op[(1,3)]{qq}$ are strongly constrained by CP violation in $K^0$-$\bar K^0$
and $D^0$-$\bar D^0$ mixing, precluding any visible effect in $\epe$, barring
unrealistic cancellations that are not stable under RG evolution.
Consequently, models with a heavy mediator that only couples to left-handed
quark doublets are not among the prime candidates to
explain a possible deviation from the SM in $\epe$.

In view of these constraints, the most attractive scenarios in the case of a tree-level
vector mediator are those that can generate the left-right operators
$\Op[(1,8)]{qu,\,qd}$.
This is even more so given that these operators eventually contribute to $\epe$
via matrix elements that are chirally enhanced.
As seen from \reftab{tab:models}, the only possibilities in this case
are a SM singlet $Z'$ or a heavy gluon $G'$, that have already been
explored in the literature (see e.g.~\cite{Buras:2014sba}), described
schematically by the following Lagrangian for $Z'$
\begin{align}
  \mathcal L_{Z'} &=
  \left[ \lambda_q^{ij} (\bar q_i \gamma_\mu q_j)
       + \lambda_u^{ij} (\bar u_i \gamma_\mu u_j)
       + \lambda_d^{ij} (\bar d_i \gamma_\mu d_j) \right] Z'^\mu \,,
\end{align}
and analogously for $G'$
\begin{align}
  \mathcal L_{G'} &=
  \left[ \lambda_q^{ij} (\bar q_i \gamma^\mu T^A q_j)
       + \lambda_u^{ij} (\bar u_i \gamma^\mu T^A u_j)
       + \lambda_d^{ij} (\bar d_i \gamma^\mu T^A d_j) \right] G'^A_\mu \,.
\end{align}

In the case of the operators $\Op[(1,8)]{qu}$, only two flavour index
combinations\footnote{We again omit redundant operators.}
contribute to the $\Delta S=1$ matching at $\muEW$,
namely $\op[(1,8)]{qu}{1211}$ and $\op[(1,8)]{qu}{1222}$.
Neglecting SMEFT RG effects, this corresponds to a product of one of the
real-valued couplings
$\lambda_{u}^{11}$ or $\lambda_{u}^{22}$
and the complex-valued coupling $\lambda_{q}^{12}$.
The square of the latter coupling also generates a contribution
to $\epsK$. Barring a fine-tuning of the phase to $\pi/2$,
this requires $|\lambda_{q}^{12}|/M_{Z'}$ to be below $(13\,\text{PeV})^{-1}$,
as seen from \reftab{tab:eps}. A visible effect in $\epe$ is then only
possible for a coupling $|\lambda_{u}^{11}|/M_{Z'}$ not smaller than
$(10\,\text{TeV})^{-1}$. For masses within reach of the LHC, this implies
a large cross section $pp\to jj$, and the $pp\to jj$ angular distribution
allows to constrain operators with flavour structure $(\bar u u)(\bar u u)$ even beyond
resonance production. Whether such a model remains viable
in view of these stringent bounds deserves a dedicated study.

In the case of the operators $\Op[(1,8)]{qd}$, more flavour index combinations
contribute to the $\Delta S=1$ matching at $\muEW$ as seen in
\refsec{sec:matching}, since they can also contribute via right-handed down-type
quarks and left-handed up-type quarks. Nevertheless, a contribution to
$\epsK$ is generated either by the 12-coupling\footnote{%
The only way to generate a $\Delta S=1$ operator at $\muEW$ without a
12-coupling is via the operators $\op[(1,8)]{qd}{1332}$; however, they match
onto $C_{SLR}^b$ and $\widetilde{C}_{SLR}^b$, which contribute to
$\epe$ neither directly nor indirectly, as shown in section~\ref{sec:rge}.
} to right-handed
or to left-handed down-type quarks.
Consequently, comparably stringent bounds as in the case of $\Op[(1,8)]{qu}$
apply.

We finally note that models where a vector mediator dominantly contributes to $\epe$ via
the purely right-handed four-quark operators $\Op[(1,8)]{ud}$ or $\Op[]{dd}$
are subject to similar constraints from $\Delta F=2$ and dijets,
but their contributions to $\epe$ are not chirally enhanced, as shown in
\refsec{sec:master}, such that a sizable contribution to $\epe$ is even
more difficult to attain.

%
%
\subsection{Models with modified electroweak couplings}\label{sec:WZmodels}

\begin{table}[tbp]
  \centering
  \begin{tabular}{clcccc}
  \toprule
    Spin & Rep. &
    $\mathcal{O}^{(1)}_{Hq}$ &
    $\mathcal{O}^{(3)}_{Hq}$ &
    $\mathcal{O}^{}_{Hd}$ &
    $\mathcal{O}^{}_{Hud}$
    \\
    \midrule
    \multirow{6}{*}{$\frac{1}{2}$} & $\left(3,1\right)_{\frac 23}$ &
    $\times$ & $\times$ \\
    & $\left(3,1\right)_{-\frac 13}$ &
    $\times$ & $\times$ \\
    & $\left(3,3\right)_{-\frac 13}$ &
    $\times$ & $\times$ \\
    & $\left(3,3\right)_{\frac 23}$ &
    $\times$ & $\times$ \\
    & $\left(3,2\right)_{\frac 16}$ &
    && $\times$ & $\times$ \\
    & $\left(3,2\right)_{-\frac 56}$ &
    && $\times$ & \\
    \midrule
    \multirow{3}{*}{1} & $\left(1,1\right)_{0}$ &
    $\times$ && $\times$ & \\
    & $\left(1,1\right)_{1}$ &
    &&& $\times$ \\
    & $\left(1,3\right)_{0}$ &
    & $\times$\\
    \bottomrule
  \end{tabular}
  \caption{SMEFT operators of type $\psi^2H^2D$ inducing corrections to
  $W^\pm$ and $Z^0$ couplings, generated by the tree-level mixing of SM fields
  with heavy vector-like quarks or vector fields.
  The second column gives the representation under
  $G_\text{SM} = SU(3)_c \otimes SU(2)_L \otimes U(1)_Y$.
  }
  \label{tab:models-zw}
\end{table}

Apart from a tree-level exchange of heavy scalar or vector bosons,
$\epe$ can also arise at tree level in the SMEFT from the operators of
type $\psi^2H^2D$ that induce modified couplings to the $Z^0$ and $W^\pm$ bosons.
In the broken phase of the SM, these contributions can be seen as arising
from the mixing between SM fermion or boson fields with heavy vector-like
fermions or vector bosons after electroweak symmetry breaking.
In \reftab{tab:models-zw}, we list all the possible vector-like fermion
or vector boson representations that generate any of the $\psi^2H^2D$
operators that give a matching contribution to $\Delta S=1$ at $\muEW$
\cite{deBlas:2017xtg}.

The vector-like fermion representations have already been discussed in detail
in the context of $\epe$ in \cite{Bobeth:2016llm}, with the exception of the
state $(3,1)_{2/3}$ that transforms like a right-handed up-type quark singlet.
In this case, one gets $C_{Hq}^{(1)}=-C_{Hq}^{(3)}$, such that there is no
flavour-changing $Z^0$ coupling and thus no contribution to semi-leptonic FCNCs
(cf.~\refsec{sec:sl}),
but a contribution to $\epe$ can nevertheless arise from a modified left-handed
$W^\pm$ coupling.

The three spin-1 models in \reftab{tab:models-zw} already appeared in
\reftab{tab:models}; these states can contribute both through tree-level
exchange leading to a four-quark SMEFT operator or through modified $W^\pm$ or
$Z^0$ couplings. Which contribution dominates depends on the size of the couplings.
Given the strong constraints from $\epsK$ on contributions from four-quark operators
in models with vector mediators
discussed in \refsec{sec:vector}, it is an interesting question how important
this constraint is when $\epe$ is dominantly generated through flavour-changing
$Z^0$ couplings. In the $(1,1)_0$ model, i.e.\ with a SM singlet $Z'$,
there are two relevant couplings for this discussion \cite{deBlas:2017xtg},
\begin{equation}
  \mathcal L \supset
  \left[ \lambda_q^{21}  \, (\bar q_2 \gamma_\mu q_1)
  + \lambda_H \, (H^\dagger i D_\mu H ) \right] Z^{\prime \mu}
  + \text{h.c.}
\end{equation}
Rescaling the couplings as $\Delta_i \equiv \lambda_i/m_{Z'}$, the Wilson coefficients relevant
for $\Delta S=1$ and  $\Delta S=2$ read \cite{deBlas:2017xtg}
\begin{align}
  \wc[(1)]{Hq}{12}   & = -\Delta_q \,\text{Re}\Delta_H
  \,, &
  \wc[(1)]{qq}{1212} & = -\frac{1}{2} \Delta_q^2
  \,.
\end{align}
In addition, a contribution to the Wilson coefficient of the purely bosonic
operator $\Op{H D}$ is generated,
\begin{align}
  \label{eq:CHD-matching}
  \Wc{H D} &= - 2 (\text{Re}\Delta_H)^2
  \,.
\end{align}
This Wilson coefficient is related to the electroweak $T$ parameter as
\begin{equation}
  T = - 2\pi v^2 \frac{g^2+g^{\prime 2}}{g^2g^{\prime 2}} \,\Wc{H D} \,.
\end{equation}
This allows to write the magnitude of the BSM effect in $\epe$ induced by $\Wc[(1)]{Hq}$
in terms of the shifts in $\epsK$ and the $T$ parameter as
\begin{equation}
  10^3\left|\frac{\varepsilon'}{\varepsilon}\right|_{\text{BSM}}
  \approx
  0.1 \, \left|
  \left[\frac{(\epsK)_\text{BSM}}{10^{-3}}\right]
  \left[\frac{T_\text{BSM}}{0.1}\right]
  \left[\frac{\text{Im}\Delta_q}{\text{Re}\Delta_q}\right]
  \right|^{\frac 12} \,.
\end{equation}
Given that the measurement of $\epsK$ agrees with the SM at the level of
$0.5\times 10^{-3}$ and the $T$ parameter at the level of $0.05$, barring
cancellations, this shows that the $Z^0$-mediated effect is strongly
constrained unless the phase of $\Delta_q$ is tuned close to $\pi/2$.

For the vector triplet $(1,3)_0$, the analogous contribution to the $T$
parameter is absent, so the $Z^0$-mediated contribution to $\epe$ could be sizable.

The $SU(2)_L$ singlet charged gauge boson $(1, 1)_1$ could arise as the
low-energy limit of a broken left-right symmetry (see e.g.~\cite{Alioli:2017ces}).
In this case, the contribution to $\epe$ is mediated by a right-handed $W^\pm$
coupling, such that $\epsK$ gives no constraint.

%
%
\subsection{Models with dipole operators}\label{sec:dipole-models}

The chromomagnetic dipole operators $O_{8g}^{(\prime)}$ can arise
in various BSM scenarios. While the corresponding matrix element
and thus the value of $P_i$ in our master formula is small,
the absence of model-independent constraints on this contribution
makes it nevertheless interesting.

Since the SMEFT dipole operator $\Op[]{dG}$ does not receive
tree-level matching contributions,
the dipole operators at the low-energy scale $\muLow$
can arise either from four-quark operators mixing into it
through RG evolution or from loop-induced matching contributions
at the UV scale $\muNP$.
Concerning the former effect, in sections~\ref{sec:rge} and \ref{sec:matching},
we have shown that SMEFT scalar operators of type $\Op[(1,8)]{quqd}$
can induce such a contribution. Whether this contribution is relevant
depends on the structure of the couplings (cf.\ \refsec{sec:scalar}):
\begin{itemize}
\item If they dominantly match onto the scalar $\Delta S=1$ operators with
flavour $(\bar sd)(\bar uu)$ in Class~C, these
have themselves also non-vanishing
$K\to\pi\pi$ matrix elements and contribute directly to $\epe$, such that the
indirect contribution via the dipole operator is negligible.
\item If they dominantly match onto the scalar $\Delta S=1$ operators with
flavour $(\bar sd)(\bar cc)$ in Class B, they indeed contribute to $\epe$ exclusively via the dipole Wilson coefficient at
the low-energy scale.
\item If the scalars couple dominantly to top quarks
(see e.g.\ \cite{Chen:2018vog}), these operators
do not match at tree-level onto the $\Delta S=1$ EFT (where top
quarks have already been integrated out), but RG evolution
above $\muEW$ (cf.\ \refsec{sec:matching}) will generate the SMEFT
dipole operator~$\Op{dG}$.
\end{itemize}

In models with heavy scalars (but no heavy fermions), also one-loop
matching contributions at the scale $\muNP$ exist. However, in the
SMEFT, where SM quarks are massless, these contributions
are IR-divergent by themselves. The divergence is cancelled by the
RG-induced contribution of the scalar four-quark operators $\Op[(1,8)]{quqd}$.

In models with heavy vectors but no heavy fermions, we expect that
typically four-quark operator contributions are more important than loop-induced
dipole operator contributions, again with the possible exception of
top quarks, where RG-induced effects above $\muEW$ are relevant.

In models with new heavy vector-like fermions
that couple to the Higgs doublet,
sizable contributions to the dipole operator can be generated from a diagram with a SM Higgs in the loop.
This gives an important constraint in models with partial
quark compositeness \cite{Agashe:2008uz,Gedalia:2009ws,Delaunay:2012cz,Vignaroli:2012si,Konig:2014iqa}.

Finally, there can of course also be loop contributions at $\muNP$
with only new heavy particles in the loop.
 This has been
  for example studied in MSSM \cite{Tanimoto:2016yfy, Kitahara:2016otd, Endo:2016aws,
  Crivellin:2017gks,Endo:2017ums},
  where scalar operators are usually omitted because they are suppressed by
  light-quark Yukawa couplings, although some might be $\tan\beta$ enhanced,
  whereas the one-loop contribution to the dipole operator is not
  suppressed.

%
%
%
\section{Summary}\label{sec:Summary}

We have presented
 for the first time a model-independent anatomy of the ratio $\epe$
in the context of  the $\Delta S = 1$ EFT with operators invariant
under QCD and QED and in the context of the SMEFT with the operators invariant under the full SM gauge group. This was only possible thanks to the very recent
 calculations of the $K\to\pi\pi$ matrix elements of BSM operators,
namely of the chromo-magnetic dipole operators by lattice QCD
\cite{Constantinou:2017sgv} and DQCD \cite{Buras:2018evv} and in particular through the calculation of matrix elements of all four-quark BSM operators, including scalar and tensor operators, by DQCD \cite{Aebischer:2018rrz}. Even if the
latter calculations have been performed in the chiral limit, they offer for the
first time a look into the world of BSM operators contributing to $\epe$.

Our main goal was to identify those new physics scenarios which
are probed by $\epe$
and which could help to
explain the emerging anomaly in $\epe$, which is signalled both by lattice
QCD results and results from the DQCD approach.  To this end we have derived
a master formula for  $\epe$, presented already in  \cite{Aebischer:2018quc},
which can be applied to any theory beyond the SM in which the Wilson coefficients
of all contributing operators have been calculated at the electroweak scale.
The relevant hadronic matrix elements of BSM operators are from the DQCD
approach and the SM ones from lattice QCD.

In the last three years a number of analyses, addressing the $\epe$ anomaly
in concrete models, appeared in the literature (see list at the beginning of
our paper) but they concentrated on   models in which NP entered exclusively
through modifications of the Wilson coefficients of SM operators. In particular the Wilson
coefficient of the dominant electroweak penguin operator $Q_8$ plays an important
role in this context as its hadronic matrix element is chirally enhanced and
in contrast to the QCD penguin operator $Q_6$ this contribution is not suppressed
by the factor $1/\omega \approx 22$ related to the $\Delta I=1/2$ rule. While we confirm
these findings through the analysis of models that generate operators of Class A, this
is a significant limitation if one wants to have a general view of possible BSM scenarios
responsible for the $\epe$ anomaly. In particular, in the absence of even approximate
values of hadronic matrix elements of BSM operators, no complete model-independent
analysis was possible until recently.

The recent calculations of BSM $K\to\pi\pi$ matrix elements, in particular
of those of scalar and tensor operators in \cite{Aebischer:2018rrz}, combined
with the EFT and in particular SMEFT analyses presented in our paper, widened
significantly our view on BSM contributions to $\epe$.

Our analysis has two main virtues:
\begin{itemize}
\item
It opens the road  to the analyses of $\epe$ in any theory beyond the
SM and allows with the help of the master formula in \refeq{eq:master}
\cite{Aebischer:2018quc}, with details presented here, to search very efficiently
for BSM scenarios behind the $\epe$ anomaly. In particular the values
of $P_i$ collected in \refapp{app:epe-MF} indicate which routes are more
promising than others, both in the context of the low-energy
EFT and SMEFT.
By implementing our results in the open source code \texttt{flavio} \cite{flavio},
testing specific BSM theories becomes particularly simple.
\item
Through our SMEFT analysis we were able to identify
correlations between $\epe$ and various observables that depend sensitively
on the operators involved. Here $\Delta S=2$, $\Delta C=2$ and electro-magnetic
dipole moments (EDM)
play a prominent role but also correlations with $\Delta S=1$ and $\Delta C=1$
provide valuable informations.
\end{itemize}

Our take-home messages are:
\begin{itemize}
\item
Tree-level vector exchanges, like $Z^\prime$ and $G^\prime$ contributions, discussed
already by various authors, can be responsible for the observed anomaly. In these
scenarios one has to face in general important constraints from  $\Delta S=2$ and $\Delta C=2$
transitions as well as direct searches and often some fine tuning is required. Here
the main role is played by the electroweak operator $Q_8$ with its Wilson coefficient
significantly modified by NP.
\item
Models with tree-level exchange of heavy colourless or coloured scalars are a new avenue,
opened by the results for BSM operators from DQCD in \cite{Aebischer:2018rrz}.
In particular scalar and tensor operators, having chirally enhanced matrix elements
and consequently large coefficients $P_i$, are candidates for the explanation
of the anomaly in question. Moreover, some of these models, in contrast to
models with  tree-level $Z^\prime$ and $G^\prime$ exchanges,
are free
from  both  $\Delta S=2$ and $\Delta C=2$ constraints.
The EDM of the neutron is an important constraint for these models,
depending on the
couplings, but does not preclude a sizable NP effect in $\epe$.
\item
Models with modified $W^\pm$ or $Z^0$ couplings can induce sizable
effects in $\epe$ without appreciable constraints from semi-leptonic
decays such as $\kpn$ or $\klpll$.
In the case of a SM singlet $Z'$ mixing with the $Z^0$, sizable $Z^0$-mediated contributions
are disfavoured by electroweak precision tests.
\end{itemize}

The future of $\epe$ in the SM and in the context of searches for NP will depend
on how accurately it can be calculated. This requires improved lattice calculations
not only of the matrix elements of SM operators but also of the BSM ones, which
are known presently only from the DQCD approach in the chiral limit. It is also
hoped that lattice QCD will be able to take into account  isospin breaking corrections
and that other lattice collaborations will attempt to calculate
hadronic matrix elements of all relevant operators. In this context we hope
that the new analysis of the RBC-UKQCD collaboration with improved matrix elements
to be expected this year will shed new light on the hinted anomaly. Such future updates
can be easily accounted for by the supplementary details on the master formula in
\refapp{app:epe-MF}.

On the short-distance side the NNLO results for QCD penguins should be available soon
\cite{Cerda-Sevilla:2016yzo}. The dominant NNLO corrections to electroweak penguins
have been calculated almost 20 years ago \cite{Buras:1999st} and, as we have pointed
out, play a significant role in removing the scale uncertainty in $m_t(\mu)$ and
the uncertainty due to
renormalization scheme dependence. Moreover, as we have seen, its inclusion increases
the size of the $\epe$ anomaly. With  present technology   a complete NNLO calculation, using the results
in \cite{Gorbahn:2004my}, should
be feasible in a not too distant future. As far as BSM operators are concerned,
a NLO analysis of their Wilson coefficients is in progress, but its importance
is not as high as of hadronic matrix elements due to significant additional
parametric uncertainties residing  in any  NP model.
In any case, in the coming years the ratio $\epe$ is expected to play a significant
role in the search for NP. In this respect, the results presented here will be
helpful in disentangling potential models of new CP violating sources beyond the
SM as well as constraining the magnitude of their effects.

\section*{Acknowledgments}

We would like to thank Jean-Marc G{\'e}rard for discussions and Aneesh Manohar
for clarifying communications.
This work was supported by the DFG cluster of excellence ``Origin and
Structure of the Universe''.

%
%

\appendix

%
%
%
\section{$\Delta S = 1$ EFT operators}\label{app:fierz}

In full generality, the $\Delta S = 1$ dimension-six effective Hamiltonian with
$\Nf$ active quark flavours,
\begin{align}
  \label{eq:DS1-Hamiltonian-5}
  \mathcal{H}_{\Delta S = 1}^{(\Nf)} &
  = -  \sum_i C_i \, O_i \,,
\end{align}
contains three classes of operators relevant to $K\to\pi\pi$ decays:

\paragraph{four-quark operators}

\begin{align}
  \label{eq:DS1-psi4-col1}
  O_{XAB}^q &
  = (\bar s^i \Gamma_X P_A d^i) (\bar q^j \Gamma_X P_B q^j) \,,
\\
  \label{eq:DS1-psi4-col8}
  \widetilde{O}_{XAB}^q &
  = (\bar s^i \Gamma_X P_A d^j) (\bar q^j \Gamma_X P_B q^i) \,,
\end{align}

\paragraph{electro- and chromo-magnetic dipole operators}

\begin{align}
  \label{eq:DS1-dipole-QED}
  O_{7\gamma}^{(\prime)} &
  = m_s(\bar s \sigma^{\mu\nu} P_{L(R)} d) F_{\mu\nu} \,,
\\
  O_{8g}^{(\prime)}      &
  = m_s(\bar s \sigma^{\mu\nu} T^A P_{L(R)} d) G^A_{\mu\nu} \,,
\end{align}

\paragraph{semi-leptonic operators}

\begin{align}
  O_{XAB}^\ell &
  = (\bar s\, \Gamma_X P_A d) (\bar \ell\, \Gamma_X P_B \ell) \,.
\end{align}
Here $i,j$ are colour indices, $A,B=L,R$, and $X=S,V,T$ with $\Gamma_S=1$,
$\Gamma_V=\gamma^\mu$, $\Gamma_T=\sigma^{\mu\nu}$.
The semi-leptonic operators can contribute to $\epe$ only via QED RG mixing and
we neglect them throughout.
Likewise, we neglect the electro-magnetic dipole operators~$O_{7\gamma}^{(\prime)}$.
This is justified because the electro- and chromo-magnetic
dipole operators mix under QCD and therefore UV complete models always generate
both operators with a suppression of $\alpha_e/\alpha_s$ for the electro-magnetic
dipole operator in $\epe$ with respect to the chromo-magnetic one.

The number of $\Delta S=1$ four-quark operators is sizable. For $N_f = 5$,
there are $10 + 10'$ (the prime denotes the number of chirality-flipped operators)
linearly independent operators for each $q=u,c,b$:
\begin{align}
  O_{VLL}^q \,,\qquad
  O_{VLR}^q \,,\qquad
  O_{SLR}^q \,,\qquad
  O_{TLL}^q \,,\qquad
\end{align}
as well as their colour-flipped ($\widetilde{O}$) and chirality-flipped ($L \leftrightarrow R$) counterparts.
For $q = d, s$, Fierz symmetry allows to eliminate half of them, leaving only $5 + 5'$ linearly independent ones.
As our $\Delta S = 1$ \textit{reference basis} we choose to eliminate $\widetilde{O}_{i}^{d,s}$ through the relations
\begin{equation}
\begin{aligned}
  \widetilde{O}_{VLL}^{d,s} &= O_{VLL}^{d,s} \,, & &&
\\
  \widetilde{O}_{VLR}^{d} &= -2 \, O_{SRL}^{d} \,, & &&
  \widetilde{O}_{SLR}^{d} &= -\frac{1}{2} O_{VRL}^{d} \,,
\\
  \widetilde{O}_{VLR}^{s} &= -2 \, O_{SLR}^{s} \,, & &&
  \widetilde{O}_{SLR}^{s} &= -\frac{1}{2} O_{VLR}^{s} \,,
\\
  \widetilde{O}_{SLL}^{d,s} &= -\frac{1}{2} O_{SLL}^{d,s} - \frac{1}{8} O_{TLL}^{d,s} \,, & &&
  \widetilde{O}_{TLL}^{d,s} &= -6\, O_{SLL}^{d,s} + \frac{1}{2} O_{TLL}^{d,s} \,,
\end{aligned}
\end{equation}
and likewise for their chirality-flipped counterparts. Hence in total there are
$40 + 40'$ linearly independent four-quark operators in $\mathcal{H}_{\Delta S = 1}^{(5)}$,
$30 + 30'$ in $\mathcal{H}_{\Delta S = 1}^{(4)}$, and $20 + 20'$ in
$\mathcal{H}_{\Delta S = 1}^{(3)}$.

We note that this reference basis coincides with the ``flavio'' basis
defined in the Wilson coefficient exchange format (WCxf) \cite{Aebischer:2017ugx}
and used in the \texttt{flavio}~\cite{flavio} and \texttt{wilson}~\cite{Aebischer:2018bkb}
packages
up to two differences,
\begin{itemize}
  \item the normalization of the operators differs,
  \item the operators in the ``flavio'' basis have the flavour structure $(\bar ds)$
  rather than $(\bar sd)$.
\end{itemize}
The complete basis can be inspected on the WCxf web site \cite{flavio-wcxf}.

%
%
%
\section{$K\to \pi\pi$ matrix elements}\label{app:Kpipi-ME}

The $K\to \pi\pi$ matrix elements $\langle O_i \rangle_I$, see \refeq{eq:def-Kpipi-ME},
of the operators $O_i$ in the $\Delta S=1$ effective Hamiltonian
are a crucial input to the prediction of $\epe$ in the SM and beyond.
In this appendix we count the number of irreducible matrix elements (i.e.\
which cannot be related to other matrix elements by exact or nearly exact
symmetries like parity and isospin) and relate the matrix elements in our
operator basis to the traditional SM operator basis.

As discussed in \refsec{sec:epesm}, the $\Delta S=1$ effective Hamiltonian
with three active quark flavours~\refeq{eq:DS1-Hamiltonian} contains 40 four-quark
operators, half of which are related to the other ones by parity, leaving at most
20 irreducible matrix elements for each of the two isospin amplitudes. Since the
operators with flavour content $(\bar sd)(\bar ss)$ are expected to be strongly
suppressed and we neglect them, this number reduces to the 15 matrix elements
\begin{equation}
  \langle \widetilde{O}_{XLB}^u \rangle_I\,,\qquad
  \langle O_{XLB}^u \rangle_I\,,\qquad
  \langle O_{XLB}^d \rangle_I\,.
\end{equation}
where $XLB=VLL$, $VLR$, $SLL$, $SLR$, or $TLL$
(note that the operators $\widetilde{O}_{XLB}^d$ are Fierz-redundant).
In addition, isospin can be used to show that\footnote{%
In the second line, the Fierz-redundant operators $\widetilde{O}_{XLB}^d$
are used for the sake of notational brevity.}
\begin{equation}
  \label{eq:isorel}
\begin{aligned}
  \langle {O}_{XLB}^u \rangle_2 + \langle {O}_{XLB}^d \rangle_2 & = 0 \,, &
\\
  \langle \widetilde{O}_{XLB}^u \rangle_2 + \langle \widetilde{O}_{XLB}^d \rangle_2 & = 0 \,.
\end{aligned}
\end{equation}
These 10 relations allow to remove 10 of the 15 $I=2$ matrix elements.
In summary, assuming strong isospin symmetry, there are in total
15 irreducible matrix elements for $I=0$
and 5 for $I=2$.
Of these, 7 and 3 are relevant in the SM, respectively.

In terms of the traditional SM operator basis \cite{Buras:1993dy},
the matrix elements of operators in our basis can be written as
\paragraph{Class A}
\begin{equation}
\begin{aligned}
  \langle O_{VLL}^u \rangle_I &
  = \frac{1}{12} \langle Q_3 \rangle_I + \frac{1}{6} \langle Q_9 \rangle_I \,,
&&&
  \langle O_{VLR}^u \rangle_I &
  = \frac{1}{12} \langle Q_5 \rangle_I + \frac{1}{6} \langle Q_7 \rangle_I \,,
\\
  \langle \widetilde{O}_{VLL}^u \rangle_I &
  = - \frac{1}{6} \langle Q_3 \rangle_I + \frac{1}{6} \langle Q_9 \rangle_I
    + \frac{1}{4} \langle Q_4 \rangle_I\,,
&&&
  \langle \widetilde{O}_{VLR}^u \rangle_I &
  = \frac{1}{12} \langle Q_6 \rangle_I + \frac{1}{6} \langle Q_8 \rangle_I \,,
\\
  \langle O_{VLL}^d \rangle_I &
  = \frac{1}{6} \langle Q_3 \rangle_I - \frac{1}{6} \langle Q_9 \rangle_I \,,
&&&
  \langle O_{VLR}^d \rangle_I &
  = \frac{1}{6} \langle Q_5 \rangle_I - \frac{1}{6} \langle Q_7 \rangle_I \,,
\\
&&&&
  \langle O_{SRL}^d \rangle_I &
  = -\frac{1}{12}\langle Q_6 \rangle_I + \frac{1}{12}\langle Q_8 \rangle_I \,.
\end{aligned}
\end{equation}
The isospin relations \eqref{eq:isorel} in this case simply imply
the vanishing of $I=2$ matrix elements of QCD penguin operators,
$\langle Q_{3,4,5,6} \rangle_2=0$.

For the remaining 10 irreducible matrix elements, we use the results
from the so-called ``SD-basis'' in tables 4 and 5 of \cite{Aebischer:2018rrz}.
They are related to the matrix elements of operators in our basis
as follows:
\paragraph{Class C}
\begin{equation}
\begin{aligned}
  \langle O_{SLL}^u \rangle_I &
  = \langle Q_2^{ \textup{SLL},u} \rangle_I \,,
&&&
  \langle O_{TLL}^u \rangle_I &
  = - \langle Q_4^{ \textup{SLL},u} \rangle_I \,,
\\
  \langle \widetilde{O}_{SLL}^u \rangle_I &
  = \langle Q_1^{ \textup{SLL},u} \rangle_I \,,
&&&
  \langle \widetilde{O}_{TLL}^u \rangle_I &
  = - \langle Q_3^{ \textup{SLL},u} \rangle_I \,.
\end{aligned}
\label{eq:ME-C}
\end{equation}
\paragraph{Class D}
\begin{equation}
\begin{aligned}
  \langle O_{SLL}^d \rangle_I &
  = \langle Q_2^{ \textup{SLL},d} \rangle_I \,,
&&&
  \langle O_{TLL}^d \rangle_I &
  = \langle Q_1^{ \textup{SLL},d} \rangle_I \,.
\end{aligned}
\label{eq:ME-D}
\end{equation}
\paragraph{Class E}
\begin{equation}
\begin{aligned}
  \langle O_{SLR}^u \rangle_I &
  = \langle Q_2^{ \textup{SLR},u} \rangle_I \,,
&&&
  \langle \widetilde{O}_{SLR}^u \rangle_I &
  = \langle Q_1^{ \textup{SLR},u} \rangle_I \,.
\end{aligned}
\label{eq:ME-E}
\end{equation}
The isospin relations \eqref{eq:isorel} allow to eliminate
the $I=2$ matrix elements in \eqref{eq:ME-C} and \eqref{eq:ME-E}.

In \reftab{tab:me-values}, we show the numerical values of the
$K\to \pi\pi$ matrix elements of all operators entering our analysis.
The ones of the SM operators are obtained from lattice QCD with RG
evolution to $\muLow = 1.3$~GeV used in our numerical analysis.

\begin{table}
\centering
\renewcommand{\arraystretch}{1.2}
\begin{tabular}{crrc}
  \toprule
  $Q_i$   & $\langle Q_i \rangle_0$                &  $\langle Q_i \rangle_2$  & Ref.
\\
\midrule
  $Q_3$     & $-0.0399(652)(118)$  & $0$              & \cite{Bai:2015nea}
\\
  $Q_4$     & $0.267(93)(65)$      & $0$              & \cite{Bai:2015nea}
\\
  $Q_5$     & $-0.179(48)(46)$     & $0$              & \cite{Bai:2015nea}
\\
  $Q_6$     & $-0.339(97)(91)$     & $0$              & \cite{Bai:2015nea}
\\
\midrule
  $Q_7$     & $0.155(37)(53)$      & $0.1220(52)(71)$ & \cite{Bai:2015nea, Blum:2015ywa}
\\
  $Q_8$     & $1.54(6)(41)$        & $0.838(28)(31)$  & \cite{Bai:2015nea, Blum:2015ywa}
\\
  $Q_9$     & $-0.197(54)(49)$     & $0.0162(3)(6)$   & \cite{Bai:2015nea, Blum:2015ywa}
\\
\midrule
$Q_1^{\text{SLL},u}$    & $-0.005(1)$ & $[-0.003]$       & \cite{Aebischer:2018rrz}
\\
$Q_2^{\text{SLL},u}$    & $-0.044(9)$ & $[-0.031]$       & \cite{Aebischer:2018rrz}
\\
$Q_3^{\text{SLL},u}$    & $-0.371(74)$ & $[-0.262]$       & \cite{Aebischer:2018rrz}
\\
$Q_4^{\text{SLL},u}$    & $-0.214(43)$ & $[-0.151]$       & \cite{Aebischer:2018rrz}
\\
  $Q_1^{\text{SLL},d}$  & $0.0070(14)$  & $-0.002$         & \cite{Aebischer:2018rrz}
\\
  $Q_2^{\text{SLL},d}$  & $-0.088(18)$ & $0.031$          & \cite{Aebischer:2018rrz}
\\
  $Q_1^{\text{SLR},u}$  & $-0.015(3)$ & $[0.003]$        & \cite{Aebischer:2018rrz}
\\
  $Q_2^{\text{SLR},u}$  & $-0.141(28)$ & $[0.050]$        & \cite{Aebischer:2018rrz}
\\
\midrule
  $O_{8g}$              & $-0.013(4)$ & $0$              & \cite{Buras:2018evv, Constantinou:2017sgv}
\\
\bottomrule
\end{tabular}
\renewcommand{\arraystretch}{1.0}
\caption{Numerical values of $K\to\pi\pi$ hadronic matrix elements used in our
  analysis. The matrix elements of the operators in the traditional SM basis
  $Q_{3\ldots9}$ are based on lattice QCD \cite{Bai:2015nea, Blum:2015ywa},
  the ones of the BSM operators and the chromo-magnetic dipole operator on DQCD
  \cite{Buras:2018evv, Aebischer:2018rrz}.
  All matrix elements are given in the $\overline{\text{MS}}$ scheme at
  $\muLow = 1.3$~GeV and in units of GeV$^3$. The normalization convention is
  chosen as $h=1$ (at variance with refs.~\cite{Bai:2015nea, Blum:2015ywa}).
  The values in square brackets are not needed since they can be expressed
  in terms of the others by isospin and Fierz relations.
  Note that the chromo-magnetic matrix element refers to our convention,
  see \refeq{eq:DS1-dipole-QCD}.
}
\label{tab:me-values}
\end{table}

%
%
\section{$\epe$ Master formula for new physics}
\label{app:epe-MF}

For the convenience of the reader, in this appendix we provide the details to
the semi-numerical master formula \refeq{eq:master} for the BSM contributions
to $\epe$ in terms of the $\Delta S=1$ Wilson coefficients at the scale
$\muEW=160\,\text{GeV}$ and the matrix elements. We reiterate that we perform
the RG evolution of NP Wilson coefficients only at the one-loop level in QCD
and QED, so we do not take into account contributions that only arise at two-loop
level. The numerical values of the $p_{ij}^{(I)}$ and $P_i$ corresponding
to the five classes of operators introduced in \refsec{sec:rge}
are listed in the following tables.
\begin{itemize}
\item \refTab{tab:metab_A} contains the contributions from the Wilson coefficients
  from Class A that multiply SM matrix elements only.
\item \refTab{tab:metab_B} contains the contributions from the Wilson coefficients
  from Class B that only enter via RG mixing into the chromo-magnetic dipole
  operator.
\item \refTab{tab:metab_C} contains the contributions from the RLRL type operators
  of Class C with flavour structure $(\bar sd)(\bar uu)$ that contribute via
  BSM matrix elements or the chromo-magnetic dipole matrix element.
\item \refTab{tab:metab_D} contains the contributions from the RLRL type operators
  from Class D with flavour structure $(\bar sd)(\bar dd)$ that contribute via
  BSM matrix elements or the chromo-magnetic dipole matrix element.
\item \refTab{tab:metab_E} contains the contributions from the RLLR type operators
  from Class E with flavour structure $(\bar sd)(\bar uu)$ that contribute via
  matrix elements of SM operators $Q_{7,8}$ and BSM matrix elements.
\end{itemize}
Besides the $p_{ij}^{(I)}$ and $P_i$, we provide in the last column of each table the suppression scale $\Lambda$ that would generate $(\epe)_\text{BSM}=10^{-3}$ for $C_i=1/\Lambda^2$.

In these tables we restrict ourselves to listing the Wilson coefficients $C_{XAB}^q$
with $A=L$ since parity invariance of QED and QCD implies that the $p_{ij}^{(I)}$
are symmetric under the interchange of all $L$ and $R$. However, the $K\to\pi\pi$
matrix elements flip their sign \cite{Kagan:2004ia}
\begin{align}
  \label{eq:ME-parity-trafo}
  \langle (\pi\pi)_I | O_{i} | K \rangle &
  = - \langle (\pi\pi)_I | [O_i]_{L\leftrightarrow R} | K \rangle .
\end{align}
In the master formula \refeq{eq:master}, this is accounted for by the
relative sign between the primed and unprimed Wilson coefficients.

The Wilson coefficients $C_{SLR}^{b,c}$ and $\widetilde{C}_{SLR}^{b,c}$
do not contribute at all at the level considered, since they do not mix at
one-loop level into any of the operators with non-vanishing $K\to\pi\pi$
matrix element.

\begin{table}[tbp]
  \centering
  \resizebox{\textwidth}{!}{
    \input{metab_A_all.tex}
  }
  \caption{Coefficients $p_{ij}^{(I)}$ and $P_i$ of the master formula
  \eqref{eq:master}--\eqref{eq:master2} as well as suppression scale $\Lambda$ (last column)
  from Wilson coefficients of operators in Class A multiplying SM matrix elements only.
  Empty entries correspond to coefficients with absolute value $<0.05$.}
  \label{tab:metab_A}
\end{table}

\begin{table}[tbp]
  \centering
  \input{metab_B_all.tex}
  \caption{Coefficients $p_{ij}^{(I)}$ and $P_i$ of the master formula
  \eqref{eq:master}--\eqref{eq:master2} as well as suppression scale $\Lambda$ (last column)
  from Wilson coefficients of Class B only entering via RG mixing into the chromo-magnetic
  dipole operator.}
  \label{tab:metab_B}
\end{table}

\begin{table}[tbp]
  \centering
  \resizebox{\textwidth}{!}{
    \input{metab_C_all.tex}
  }
  \caption{Coefficients $p_{ij}^{(I)}$ and $P_i$ of the master formula
  \eqref{eq:master}--\eqref{eq:master2} as well as suppression scale $\Lambda$ (last column)
  from RLRL type operators of Class C with flavour structure $(\bar sd)(\bar uu)$.}
  \label{tab:metab_C}
\end{table}

\begin{table}[tbp]
  \centering
  \input{metab_D_all.tex}
  \caption{Coefficients $p_{ij}^{(I)}$ and $P_i$ of the master formula
  \eqref{eq:master}--\eqref{eq:master2} as well as suppression scale $\Lambda$ (last column)
  from RLRL type operators of Class D with flavour structure $(\bar sd)(\bar dd)$.}
  \label{tab:metab_D}
\end{table}

\begin{table}[tbp]
  \centering
  \input{metab_E_all.tex}
  \caption{Coefficients $p_{ij}^{(I)}$ and $P_i$ of the master formula
  \eqref{eq:master}--\eqref{eq:master2} as well as suppression scale $\Lambda$ (last column)
  from RLLR type operators of Class E with flavour structure $(\bar sd)(\bar uu)$.
  Empty entries correspond to vanishing coefficients.}
  \label{tab:metab_E}
\end{table}

%
%
%
\section{SMEFT operators}
\label{app:SMEFT-operators}

In general, the following classes of SMEFT operators can contribute to the matching
onto the $\Delta S=1$ EFT at $\muEW$:
\begin{itemize}
\item The $\psi^2 H X$ dipole operators $\Op{dB},\; \Op{dW},\; \Op{dG}$.
\item The $\psi^4$ non-leptonic operators
  $\Op{dd},\; \Op[(1,8)]{ud},\; \Op[(1,8)]{qu},\; \Op[(1,8)]{qd},\;
    \Op[(1,3)]{qq},\; \Op[(1,8)]{quqd}$.
\item Contributions from modified $W^\pm$ and $Z^0$ couplings are generated by
  $\psi^2 H^2 D$ operators $\Op[(1,3)]{H q}$ and $\Op{H d}$ that
  mediate both non- and semi-leptonic transitions. The $\psi^2 H^3$ operator
  $\Op{dH}$ parametrizes modified $h^0$ couplings and contributes via
  tree-level $h^0$ exchange, but for light quark- and lepton-Yukawa couplings
  such exchange counts as a dimension-eight contribution \cite{Jenkins:2017jig}.
\item The $\psi^4$ semi-leptonic operators
  $\Op[(1,3)]{\ell q},\; \Op{qe},\; \Op{\ell d},\; \Op{ed},\; \Op{\ell edq}$.
\end{itemize}
We follow the SMEFT conventions of ref.~\cite{Grzadkowski:2010es} and provide
the definitions of the above operators in tables~\ref{tab:SMEFT-4Fops}
and~\ref{tab:SMEFT-magnops}, as well as those operators that mix into Classes A--C
operators listed in tables~\ref{tab::OpMixing} and~\ref{tab::OpMixinggauge}.

\begin{table}[tbp]
\centering
\renewcommand{\arraystretch}{1.5}
\begin{tabular}{cc|cc}
\toprule
\multicolumn{2}{c|}{$(\bar{L}R)(\bar{R}L)$ or $(\bar{L}R)(\bar{L}R)$}&
\multicolumn{2}{c}{$(\bar{L}L)(\bar{L}L)$}
\\
\hline
  $\Op{\ell edq}$    & $(\bar{\ell }_i^a e_j) (\bar{d}_k q_l^a)$
& $\Op[(1)]{qq}$     & $(\bar q_i \gamma_\mu q_j)(\bar q_k \gamma^\mu q_l)$
\\
  $\Op[(1)]{quqd}$   & $(\bar q^a_i u_j) \varepsilon_{ab} (\bar q^b_k d_l)$
& $\Op[(1)]{\ell q}$ & $(\bar \ell_i \gamma_\mu \ell_j)(\bar q_k \gamma^\mu q_l)$
\\
  $\Op[(8)]{quqd}$   & $(\bar q^a_i T^{\mysmall A} u_j) \varepsilon_{ab} (\bar q^b_k T^{\mysmall A} d_l)$
& $\Op[(3)]{qq}$     & $(\bar q_i \gamma_\mu \tau^I q_j)(\bar q_k \gamma^\mu \tau^I q_l)$
\\
\cline{1-2}
  \multicolumn{2}{c|}{$(\bar{L}L)(\bar{R}R)$}
& $\Op[(3)]{\ell q}$ & $(\bar \ell_i \gamma_\mu \tau^I \ell_j)(\bar q_k \gamma^\mu \tau^I q_l)$
\\
\cline{1-2}
  $\Op{\ell u}$      & $(\bar \ell_i \gamma_\mu \ell_j)(\bar u_k \gamma^\mu u_l)$
& &
\\
\cline{3-4}
  $\Op{\ell d}$      & $(\bar \ell_i \gamma_\mu \ell_j)(\bar d_k \gamma^\mu d_l)$
& \multicolumn{2}{c}{$(\bar{R}R)(\bar{R}R)$}
\\
\cline{3-4}
  $\Op{qe}$          & $(\bar q_i \gamma_\mu q_j)(\bar e_k \gamma^\mu e_l)$
& $\Op{dd}$          & $(\bar d_i \gamma_\mu d_j)(\bar d_k \gamma^\mu d_l)$
\\
  $\Op[(1)]{qu}$     & $(\bar q_i \gamma_\mu q_j)(\bar u_k \gamma^\mu u_l)$
& $\Op{eu}$          & $(\bar e_i \gamma_\mu e_j)(\bar u_k \gamma^\mu u_l)$
\\
  $\Op[(1)]{qd}$     & $(\bar q_i \gamma_\mu q_j)(\bar d_k \gamma^\mu d_l)$
& $\Op{ed}$          & $(\bar e_i \gamma_\mu e_j)(\bar d_k \gamma^\mu d_l)$
\\
  $\Op[(8)]{qu}$     & $(\bar q_i \gamma_\mu T^{\mysmall A} q_j)(\bar u_k \gamma^\mu T^{\mysmall A} u_l)$
& $\Op[(1)]{ud}$     & $(\bar u_i \gamma_\mu u_j)(\bar d_k \gamma^\mu d_l)$
\\
  $\Op[(8)]{qd}$     & $(\bar q_i \gamma_\mu T^{\mysmall A} q_j)(\bar d_k \gamma^\mu T^{\mysmall A} d_l)$
& $\Op[(8)]{ud}$     & $(\bar u_i \gamma_\mu T^{\mysmall A} u_j)(\bar d_k \gamma^\mu T^{\mysmall A} d_l)$
\\
\bottomrule
\end{tabular}
\renewcommand{\arraystretch}{1.0}
  \caption{List of the dimension-six four-fermion ($\psi^4$) operators in SMEFT
  that contribute to $s \to d$ transitions at tree level or via mixing. Flavour
  indices on the quark and lepton fields are $ijkl$.}
  \label{tab:SMEFT-4Fops}
\end{table}
\begin{table}[tbp]
\centering
\renewcommand{\arraystretch}{1.5}
\begin{tabular}{cc|cc}
\toprule
& $\psi^2 X H$
&
& $\psi^2 H^2 D$
\\
\hline
  $\Op{uB}$
& $(\bar{q}_i \sigma^{\mu\nu} u_j) \widetilde{H} B_{\mu\nu}$
& $\Op[(1)]{Hq}$
& $(H^\dagger i \overleftrightarrow{D}_{\!\!\mu} H) (\bar{q}_i \gamma^\mu q_j)$
\\
  $\Op{dB}$
& $(\bar{q}_i \sigma^{\mu\nu} d_j) H B_{\mu\nu}$
& $\Op[(3)]{Hq}$
& $(H^\dagger i \overleftrightarrow{D}^I_{\!\!\mu} H) (\bar{q}_i \tau^I \gamma^\mu q_j)$
\\
  $\Op{uW}$
& $(\bar{q}_i \sigma^{\mu\nu} u_j) \tau^{\mysmall I} \widetilde{H} W^{\mysmall I}_{\mu\nu}$
& $\Op{Hu}$
& $(H^\dagger i \overleftrightarrow{D}_{\!\!\mu} H) (\bar{u}_i \gamma^\mu u_j)$
\\
  $\Op{dW}$
& $(\bar{q}_i \sigma^{\mu\nu} d_j) \tau^{\mysmall I} H W^{\mysmall I}_{\mu\nu}$
& $\Op{Hd}$
& $(H^\dagger i \overleftrightarrow{D}_{\!\!\mu} H) (\bar{d}_i \gamma^\mu d_j)$
\\
  $\Op{uG}$
& $(\bar{q}_i \sigma^{\mu\nu} T^{\mysmall A} u_j) \widetilde{H} G^{\mysmall A}_{\mu\nu}$
& $\Op{Hud}$
& $(\widetilde{H}^\dagger i {D}_{\!\mu} H) (\bar{u}_i \gamma^\mu d_j)$
\\
  $\Op{dG}$
& $(\bar{q}_i \sigma^{\mu\nu} T^{\mysmall A} d_j) H G^{\mysmall A}_{\mu\nu}$
&
&
\\
\bottomrule
\end{tabular}
\renewcommand{\arraystretch}{1.0}
\caption{Dimension-six electro- and chromo-magnetic dipole ($\psi^2 H X$)
  and $\psi^2 H^2 D$ operators in SMEFT.}
  \label{tab:SMEFT-magnops}
\end{table}

\section{SMEFT matching conditions}
\label{app:SMEFT-matching}

\subsection{Four-quark operators}
\label{sec:4q}

The tree-level matching of the SMEFT four-quark operators yields the following
non-vanishing matching conditions  for
the Wilson coefficients in the $\Delta S = 1$~EFT~\refeq{eq:DS1-Hamiltonian-5},
\begin{align}
  C_{VLL}^{d_i} & = \wc[(1)]{qq}{21ii} + \wc[(3)]{qq}{21ii} \,, &
  \widetilde{C}_{VLL}^{d_i} & = \wc[(1)]{qq}{2ii1} + \wc[(3)]{qq}{2ii1}
  \,,\label{eq:match1}
\end{align}
\begin{align}
  C_{VLL}^{u_i} & = \sum_{jk} V_{ij}^{} V_{ik}^\ast \left(
    \wc[(1)]{qq}{21jk} + \wc[(1)]{qq}{jk21}
  - \wc[(3)]{qq}{21jk} - \wc[(3)]{qq}{jk21} \right) \,,
\\
  \widetilde{C}_{VLL}^{u_i} & = 2\,\sum_{jk} V_{ij}^{} V_{ik}^\ast  \left(
     \wc[(3)]{qq}{j12k} + \wc[(3)]{qq}{2kj1} \right) \,,
\end{align}
\begin{align}
  C_{VRR}^{d_i} & = \wc[]{dd}{21ii} \,, &
  \widetilde{C}_{VRR}^{d_i} & = \wc[]{dd}{2ii1} \,,
\\
  C_{VRR}^{u_i} & = \wc[(1)]{ud}{ii21} - \frac{1}{6} \wc[(8)]{ud}{ii21} \,, &
  \widetilde{C}_{VRR}^{u_i} & =  \frac{1}{2} \wc[(8)]{ud}{ii21} \,,
\end{align}
\begin{align}
  C_{VLR}^{u_i} & = \wc[(1)]{qu}{21ii} - \frac{1}{6} \wc[(8)]{qu}{21ii} \,, &
  \widetilde{C}_{VLR}^{u_i} & = \frac{1}{2} \wc[(8)]{qu}{21ii} \,,
\\
  C_{VRL}^{u_i} & = \sum_{jk} V_{ij}^{} V_{ik}^* \left (
    \wc[(1)]{qd}{jk21} - \frac{1}{6} \wc[(8)]{qd}{jk21} \right)\,, &
  \widetilde{C}_{VRL}^{u_i} & = \frac{1}{2} \sum_{jk} V_{ij}^{} V_{ik}^* \wc[(8)]{qd}{jk21} \,,
\\
  C_{VLR}^{d_i} & =  \wc[(1)]{qd}{21ii} - \frac{1}{6} \wc[(8)]{qd}{21ii} \,, &
  \widetilde{C}_{VLR}^{d_i} & = \frac{1}{2} \wc[(8)]{qd}{21ii} \,,
\\
  C_{VRL}^{d_i} & = \wc[(1)]{qd}{ii21} - \frac{1}{6} \wc[(8)]{qd}{ii21} \,, &
  \widetilde{C}_{VRL}^{d_i} & = \frac{1}{2} \wc[(8)]{qd}{ii21} \,,
\end{align}
\begin{align}
  C_{SRL}^{d_i} & = -\wc[(8)]{qd}{2ii1} \,, &
  \widetilde{C}_{SRL}^{d_i} & = -2 \wc[(1)]{qd}{2ii1} + \frac{1}{3} \wc[(8)]{qd}{2ii1} \,,
\\
  C_{SLR}^{d_i} & = -\wc[(8)]{qd}{i12i} \,, &
  \widetilde{C}_{SLR}^{d_i} & = -2 \wc[(1)]{qd}{i12i} + \frac{1}{3}\wc[(8)]{qd}{i12i} \,,
\end{align}
\begin{align}
  C_{SRR}^{u_i} & = \sum_j V_{ij} \left(
      \wc[(1)]{quqd}{ji21} + \frac{1}{4}\wc[(8)]{quqd}{2ij1}
    - \frac{1}{6} \wc[(8)]{quqd}{ji21} \right) \,,
\\
  \widetilde{C}_{SRR}^{u_i} & = \sum_j V_{ij} \left(
    \frac{1}{2}\wc[(1)]{quqd}{2ij1} - \frac{1}{12}\wc[(8)]{quqd}{2ij1}
  + \frac{1}{2} \wc[(8)]{quqd}{ji21} \right) \,,
\\
  C_{SLL}^{u_i} & = \sum_j  V_{ij}^* \left(
    \wc[(1)]{quqd}{ji12}^* + \frac{1}{4}\wc[(8)]{quqd}{1ij2}^*
  - \frac{1}{6} \wc[(8)]{quqd}{ji12}^* \right) \,,
\\
  \widetilde{C}_{SLL}^{u_i} & = \sum_j V_{ij}^* \left(
     \frac{1}{2}\wc[(1)]{quqd}{1ij2}^* - \frac{1}{12}\wc[(8)]{quqd}{1ij2}^*
   + \frac{1}{2} \wc[(8)]{quqd}{ji12}^* \right) \,,
\end{align}
\begin{align}
  C_{TRR}^{u_i} & = \frac{1}{16} \sum_j V_{ij} \wc[(8)]{quqd}{2ij1} \,, &
  \widetilde{C}_{TRR}^{u_i} & = \sum_j V_{ij} \left(
    \frac{1}{8} \wc[(1)]{quqd}{2ij1} - \frac{1}{48} \wc[(8)]{quqd}{2ij1}   \right) \,,
\\
  C_{TLL}^{u_i} & = \frac{1}{16}\sum_j V_{ij}^* \wc[(8)]{quqd}{1ij2}^* \,, &
  \widetilde{C}_{TLL}^{u_i} & = \sum_j V_{ij}^* \left(
    \frac{1}{8} \wc[(1)]{quqd}{1ij2}^* - \frac{1}{48} \wc[(8)]{quqd}{1ij2}^* \right) \,.
  \label{eq:matchlast}
\end{align}
where $V$ is the CKM matrix and we  have explicitly written the sum over $j$
on the right-hand side where necessary, while $i$ is not to be summed over.

\subsection{Modified $Z^0$ and $W^\pm$ couplings}
\label{sec:ZW}

In addition to the direct matching of four-quark SMEFT operators onto
four-quark $\Delta S=1$ EFT operators, the latter also receive dimension-six
matching contributions from diagrams with tree-level $Z^0$ or $W^\pm$
exchange, with one SM coupling and the other from a SMEFT $\psi^2H^2D$ operator
of modified $Z^0$ or $W^\pm$ coupling.

The $Z^0$ exchanges lead to the following additional matching contributions
to vector operators,
\begin{equation}
\begin{aligned}
  C_{VLA}^{u_i} & =  2\,\zeta_{u_A} \left[\Wc[(1)]{Hq} + \Wc[(3)]{Hq} \right]_{12}^* \,, & &&
  C_{VRA}^{u_i} & =  2\,\zeta_{u_A} \wc[]{Hd}{12}^*  \,,
\\
  C_{VLA}^{d_i} & =  2\,\zeta_{d_A} \left[\Wc[(1)]{Hq} + \Wc[(3)]{Hq} \right]_{12}^* \,, & &&
  C_{VRA}^{d_i} & =  2\,\zeta_{d_A} \wc[]{Hd}{12}^* \,,
\end{aligned}
\end{equation}
where we have written the SM $Z^0$ coupling to quarks $q=u,d$ and $A=L,R$, as
\begin{equation}
  \mathcal L_\text{SM} \supset
  \frac{g}{\cos\theta_W}\,\zeta_{q_A} \left(\bar q \gamma^\mu P_A q \right)Z^0_\mu \,,
\end{equation}
with
\begin{align}
  \zeta_{q_L} & = T_3^q - Q_q \sin^2\theta_W  \,,&
  \zeta_{q_R} & = - Q_q \sin^2\theta_W \,,
\end{align}
and the $SU(2)_L$ coupling $g$.

In the case of $W^\pm$ exchange, there are two qualitatively different contributions.
The first involves a modified $W^\pm$ coupling to left-handed quarks induced
by the operator $\Op[(3)]{Hq}$ and affects
the matching contribution of $\widetilde{C}_{VLL}^{u_i}$,
\begin{align}
  \widetilde{C}_{VLL}^{u_i} & =  -2\sum_j
  \left(
  \wc[(3)]{Hq}{j2}^* V_{id}^{} V_{ij}^*
  +
  \wc[(3)]{Hq}{1j}^*   V_{ij}^{} V_{is}^*
  \right)
  \,,
\end{align}
where again the sum over $j$ has been made explicit and $i$ is not to be
summed over. Here only the terms with  Wilson coefficients $\wc[(3)]{Hq}{kl}$ with
$kl=12,13,23$ are relevant for $\epe$,\footnote{%
Omitting coefficients that are redundant due to $\Wc[(3)]{Hq}$ being hermitian.}
because for $k=l$ the $\wc[(3)]{Hq}{kl}$ is manifestly
real-valued whereas the accompanying CKM factor is also real-valued ($k = 1$)
or has a negligible phase ($k=2$), such that
there is no contribution to $\epe$.

The second contribution originates from the $W^\pm$ coupling to right-handed quarks
induced by the operator $\Op{Hud}$. In this case the only non-vanishing
matching conditions are
\begin{align}
  \widetilde{C}_{SLR}^{u_i} & = 2\,V_{id}^{}\,\wc[]{Hud}{i2}^*
  \,,&
  \widetilde{C}_{SRL}^{u_i} & = 2\,V_{is}^* \,\wc[]{Hud}{i1}
  \,.
\end{align}
Since the operators $\widetilde{C}_{SLR,SRL}^{c}$ do not contribute
to $\epe$ at the one-loop level as discussed in \refsec{sec:rge},
only the case $i=1$ is relevant.

The effect of the right-handed $W^\pm$ coupling on $\epe$ has been discussed recently
in \cite{Cirigliano:2016yhc,Alioli:2017ces}
and of the other $\psi^2H^2D$ operators
in \cite{Bobeth:2016llm,Endo:2016tnu, Bobeth:2017xry}.

\subsection{Dipole operators}

Since we neglect the electro-magnetic dipole operators,
the only relevant matching conditions are those of
the chromo-magnetic operators, that trivially read
\begin{align}
  C_{8g} & = \frac{v}{\sqrt{2}m_s}\wc[]{dG}{12}^* \,, &
  C_{8g}' & = \frac{v}{\sqrt{2}m_s}\wc[]{dG}{21} \,, &
  \label{eq:match8g}
\end{align}
taking into account our normalization in \eqref{eq:DS1-dipole-QCD}.
Here $v\approx246$\,GeV is the Higgs vacuum expectation value.

\section{RG evolution in SMEFT}
\label{app:SMEFT-running}

In \reftab{tab::OpMixing} all operators are listed that mix through the large
top-quark Yukawa coupling into the four-quark SMEFT operators in Classes A--C. For Class A, these are
either four-quark operators or $\psi^2 H^2 D$ operators describing modified
$W^\pm$ and $Z^0$ couplings whereas in the case of Classes B--C only scalar
four-quark operators contribute. The corresponding mixing through gauge
couplings is given in \reftab{tab::OpMixinggauge}. For Class A, there are
four-quark and semi-leptonic operators and again $\psi^2 H^2 D$ operators,
whereas in the case of Classes B--C scalar four-quark and dipole operators
$\psi^2 H X$ contribute.

\begin{table}
\centering
\renewcommand{\arraystretch}{1.4}
\begin{tabular}{cll}
\toprule
Class A  & non-leptonic $\psi^4$                              & $\psi^2 H^2 D$   \\\midrule
$\Op[(1)]{qq}$         & $\Op[(1)]{qq}$,\, $\Op[(1,8)]{qu}$                 & $\Op[(1)]{Hq}$ \\
$\Op[(3)]{qq}$         & $\Op[(3)]{qq}$,\, $\Op[(8)]{qu}$                   & $\Op[(3)]{Hq}$ \\
$\Op[(1)]{qu}$         & $\Op[(1,8)]{qu}$,\, $\Op[(1,3)]{qq}$,\, $\Op{uu}$  & $\Op[(1)]{Hq}$,\, $\Op{Hu}$ \\
$\Op[(8)]{qu}$         & $\Op[(1,8)]{qu}$,\, $\Op[(1,3)]{qq}$,\, $\Op{uu}$  & $-$ \\
$\Op[(1)]{qd}$         & $\Op[(1)]{ud}$,\, $\Op[(1)]{qd}$                   & $\Op{Hd}$ \\
$\Op[(8)]{qd}$         & $\Op[(8)]{ud}$,\, $\Op[(8)]{qd}$                   & $-$ \\
$\Op[(1)]{ud}$         & $\Op[(1)]{ud}$,\, $\Op[(1)]{qd}$                   & $\Op{Hd}$ \\
$\Op[(8)]{ud}$         & $\Op[(8)]{ud}$,\, $\Op[(8)]{qd}$                   & $-$ \\
$\Op{dd}$              & $-$                                                & $-$  \\
$\Op[(1)]{Hq}$        & $\Op[(1,3)]{qq}$,\,$\Op[(1)]{qu}$
                      & $\Op[(1)]{Hq}$ \, \\
$\Op[(3)]{Hq}$        & $\Op[(1,3)]{qq}$
                      & $\Op[(3)]{Hq}$ \, \\
$\Op[]{Hd}$        & $\Op[(1)]{qd}$ ,\, $\Op[(1)]{ud}$
                      & $\Op[]{Hd}$ \,
\\[1em]
\toprule
Class B + C      & non-leptonic $\psi^4$                & $\psi^2 H X$   \\
\midrule
$\Op{dG}$        & $\Op[(1,8)]{quqd}$  & $\Op{dG}$ \\
$\Op[(1)]{quqd}$ & $\Op[(1,8)]{quqd}$  & \\
$\Op[(8)]{quqd}$ & $\Op[(8)]{quqd}$    & \\
\bottomrule
\end{tabular}
\renewcommand{\arraystretch}{1.0}
\caption{List of four-quark operators in Classes A--C that receive at one-loop in SMEFT contributions
  from non-leptonic $\psi^4$ and $\psi^2 H^2 D$ operators through mixing via the
  top-quark Yukawa coupling. Self-mixing is included.
  We have omitted $\Op[]{Hud}$ (Class E) that only mixes with itself
  through the top Yukawa coupling.}
\label{tab::OpMixing}
\end{table}

\begin{table}
\centering
\renewcommand{\arraystretch}{1.4}
\begin{tabular}{clll}
\toprule
Class A & non-leptonic $\psi^4$ & semi-leptonic $\psi^4$ & $\psi^2 H^2 D$
\\
\midrule
  $\Op[(1)]{qq}$
& $\Op[(1,3)]{qq}$, $\Op[(1,8)]{qu}$, $\Op[(1,8)]{qd}$
& $\Op[(1)]{\ell q}$, $\Op{qe}$
& $\Op[(1)]{Hq}$
\\
  $\Op[(3)]{qq}$
& $\Op[(1,3)]{qq}$, $\Op[(8)]{qu}$, $\Op[(8)]{qd}$
& $\Op[(3)]{\ell q}$
& $\Op[(3)]{Hq}$
\\
  $\Op[(1)]{qu}$
& $\Op[(1,3)]{qq}$, $\Op[(1,8)]{qu}$, $\Op[(1)]{qd}$, $\Op[(1)]{ud}$, $\Op{uu}$
& $\Op[(1)]{\ell q}$, $\Op{qe}$, $\Op{\ell u}$, $\Op{eu}$
& $\Op[(1)]{Hq}$, $\Op{Hu}$
\\
  $\Op[(8)]{qu}$
& $\Op[(1,3)]{qq}$, $\Op[(1,8)]{qu}$, $\Op[(8)]{qd}$, $\Op[(8)]{ud}$, $\Op{uu}$
& &
\\
  $\Op[(1)]{qd}$
& $\Op[(1,3)]{qq}$, $\Op[(1)]{qu}$, $\Op[(1,8)]{qd}$, $\Op[(1)]{ud}$, $\Op[]{dd}$
& $\Op[(1)]{\ell q}$, $\Op{qe}$, $\Op{\ell d}$, $\Op{ed}$
& $\Op[(1)]{Hq}$, $\Op{Hd}$
\\
$\Op[(8)]{qd}$
& $\Op[(1,3)]{qq}$, $\Op[(8)]{qu}$, $\Op[(1,8)]{qd}$, $\Op[(8)]{ud}$, $\Op{dd}$
& &
\\
  $\Op[(1)]{ud}$
& $\Op[(1,8)]{ud}$,  $\Op[(1)]{qu}$, $\Op[(1)]{qd}$, $\Op{uu}$, $\Op{dd}$
& $\Op{\ell u}$, $\Op{\ell d}$, $\Op{eu}$, $\Op{ed}$
& $\Op{Hu}$, $\Op{Hd}$
\\
  $\Op[(8)]{ud}$
& $\Op[(8)]{qu}$, $\Op[(8)]{qd}$, $\Op[(1,8)]{ud}$, $\Op{uu}$, $\Op{dd}$
& &
\\
  $\Op{dd}$
& $\Op[(1,8)]{qd}$, $\Op[(1,8)]{ud}$, $\Op{dd}$
& $\Op{\ell d}$, $\Op{ed}$
& $\Op{Hd}$ \\
$\Op[(1)]{Hq}$        & $\Op[(1,3)]{qq}$,\,$\Op[(1)]{qu}$,\,$\Op[(1)]{qd}$
                      & $\Op[(1)]{\ell q}$, $\Op{qe}$
                      & $\Op[(1)]{Hq}$ \, \\
$\Op[(3)]{Hq}$        & $\Op[(1,3)]{qq}$
                      & $\Op[(3)]{\ell q}$
                      & $\Op[(3)]{Hq}$ \, \\
$\Op[]{Hd}$        & $\Op[(1)]{qd}$ ,\, $\Op[(1)]{ud}$,\,$\Op{dd}$,\,$\Op[(1)]{ud}$
                  & $\Op{\ell d}$,\,$\Op{e d}$
                      & $\Op[]{Hd}$ \,
\\[1em]
\toprule
Class B+C  & non-leptonic $\psi^4$  & $\psi^2 H X$
\\
\midrule
  $\Op{dG}$
&
& $\Op{dG}$, $\Op{dB}$, $\Op{dW}$
&
\\
  $\Op[(1)]{quqd}$
& $\Op[(1,8)]{quqd}$
& $\Op{uG}$, $\Op{dG}$, $\Op{uB}$, $\Op{uW}$, $\Op{dB}$, $\Op{dW}$
&
\\
  $\Op[(8)]{quqd}$
& $\Op[(1,8)]{quqd}$
& $\Op{uG}$, $\Op{dG}$, $\Op{uB}$, $\Op{uW}$, $\Op{dB}$, $\Op{dW}$
& \\
\bottomrule
\end{tabular}
\renewcommand{\arraystretch}{1.0}
\caption{List of SMEFT operators that mix in SMEFT at one-loop with
  the ones in Classes A--C through gauge couplings. Self-mixing is included.
  We have omitted $\Op[]{Hud}$ (Class E) that only mixes with itself
  through gauge couplings.}
\label{tab::OpMixinggauge}
\end{table}

%
%
%

\newpage

\bibliographystyle{JHEP}
\bibliography{bibliography}

\end{document}

%% file: metab_A_all.tex
\begin{tabular}{lrrrrlllrrrcr}
\toprule
{} &  $\langle Q_3 \rangle_0$ &  $\langle Q_4 \rangle_0$ &  $\langle Q_5 \rangle_0$ &  $\langle Q_6 \rangle_0$ & $\langle Q_7 \rangle_0$ & $\langle Q_8 \rangle_0$ & $\langle Q_9 \rangle_0$ &  $\langle Q_7 \rangle_2$ &  $\langle Q_8 \rangle_2$ &  $\langle Q_9 \rangle_2$ &              $P_i$ &  $\frac{\Lambda}{\text{TeV}}$ \\
\midrule
$C_{VLL}^{u}$             &                     10.7 &                     $-7.4$ &                      &                      0.2 &                   $-0.07$ &                   $-0.04$ &                    6.32 &                      1.6 &                      0.8 &                   $-141.8$ &     $-4.3 \pm 1.0$ &                            65 \\
$C_{VLR}^{u}$             &                      0.1 &                     $-0.2$ &                      3.7 &                      3.6 &                    7.28 &                    7.73 &                   $-0.06$ &                   $-163.4$ &                   $-173.4$ &                      1.4 &   $-126 \pm 10$ &                           354 \\
$\widetilde{C}_{VLL}^{u}$ &                    $-13.6$ &                     16.4 &                      0.2 &                     $-1.2$ &                   $-0.01$ &                         &                    6.37 &                      0.3 &                      0.1 &                   $-143.0$ &      $1.5 \pm 1.7$ &                            38 \\
$\widetilde{C}_{VLR}^{u}$ &                      0.7 &                     $-1.3$ &                      0.4 &                     13.4 &                   $-0.04$ &                   30.62 &                   $-0.04$ &                      1.0 &                   $-687.4$ &                      0.9 &  $-436 \pm 35$ &                           659 \\
$C_{VLL}^{d}$             &                      6.6 &                     $-0.6$ &                      0.1 &                     $-0.9$ &                    0.04 &                    0.03 &                   $-6.26$ &                     $-0.9$ &                     $-0.6$ &                    140.6 &      $2.3 \pm 0.5$ &                            48 \\
$C_{VLR}^{d}$             &                      0.1 &                     $-0.3$ &                      7.5 &                      7.5 &                   $-7.41$ &                   $-7.84$ &                    0.03 &                    166.4 &                    176.1 &                     $-0.7$ &   $123 \pm 10$ &                           350 \\
$C_{SLR}^{d}$             &                      0.3 &                     $-0.7$ &                      0.2 &                     14.5 &                    0.01 &                  $-15.51$ &                    0.01 &                     $-0.2$ &                    348.2 &                     $-0.2$ &   $214 \pm 18$ &                           462 \\
$C_{VLL}^{s}$             &                      0.3 &                     $-0.6$ &                      0.1 &                     $-0.9$ &                    0.04 &                    0.03 &                    0.03 &                     $-0.9$ &                     $-0.6$ &                     $-0.8$ &     $-0.4 \pm 0.1$ &                            18 \\
$C_{VLR}^{s}$             &                      0.1 &                     $-0.3$ &                      0.1 &                     $-0.3$ &                    0.03 &                    0.02 &                    0.03 &                     $-0.8$ &                     $-0.4$ &                     $-0.7$ &     $-0.32 \pm 0.05$ &                            17 \\
$C_{SLR}^{s}$             &                     $-0.3$ &                      0.7 &                     $-0.2$ &                      1.0 &                   $-0.01$ &                   $-0.01$ &                   $-0.01$ &                      0.2 &                      0.2 &                      0.2 &      $0.04 \pm 0.13$ &                             6 \\
$C_{VLL}^{c}$             &                     $-0.1$ &                      0.2 &                      &                      0.2 &                   $-0.07$ &                   $-0.04$ &                   $-0.06$ &                      1.6 &                      0.8 &                      1.4 &      $0.7 \pm 0.1$ &                            25 \\
$C_{VLR}^{c}$             &                      0.1 &                     $-0.2$ &                      0.1 &                     $-0.3$ &                   $-0.07$ &                   $-0.03$ &                   $-0.06$ &                      1.6 &                      0.8 &                      1.4 &      $0.7 \pm 0.1$ &                            26 \\
$\widetilde{C}_{VLL}^{c}$ &                      0.4 &                     $-0.7$ &                      0.2 &                     $-1.2$ &                   $-0.01$ &                         &                   $-0.01$ &                      0.3 &                      0.1 &                      0.2 &      $0.2 \pm 0.2$ &                            13 \\
$\widetilde{C}_{VLR}^{c}$ &                      0.7 &                     $-1.3$ &                      0.4 &                     $-1.9$ &                   $-0.04$ &                   $-0.01$ &                   $-0.04$ &                      1.0 &                      0.2 &                      0.9 &      $0.4 \pm 0.2$ &                            20 \\
$C_{VLL}^{b}$             &                      &                     &                      &                      0.1 &                    0.02 &                    0.02 &                    0.02 &                     $-0.6$ &                     $-0.4$ &                     $-0.5$ &     $-0.30 \pm 0.03$ &                            17 \\
$C_{VLR}^{b}$             &                     &                     $-0.1$ &                     &                     $-0.1$ &                    0.02 &                    0.02 &                    0.02 &                     $-0.6$ &                     $-0.4$ &                     $-0.5$ &     $-0.28 \pm 0.03$ &                            16 \\
$\widetilde{C}_{VLL}^{b}$ &                      0.3 &                     $-0.4$ &                      0.1 &                     $-0.8$ &                         &                    0.01 &                         &                     $-0.1$ &                     $-0.2$ &                     $-0.1$ &     $-0.02 \pm 0.09$ &                             4 \\
$\widetilde{C}_{VLR}^{b}$ &                      0.4 &                     $-0.6$ &                      0.1 &                     $-1.1$ &                    0.01 &                    0.01 &                    0.01 &                     $-0.2$ &                     $-0.3$ &                     $-0.2$ &     $-0.1 \pm 0.1$ &                             8 \\
\bottomrule
\end{tabular}

%% file: metab_B_all.tex
\begin{tabular}{lrcr}
\toprule
{} &  $\langle Q_g^- \rangle_0$ &              $P_i$ &  $\frac{\Lambda}{\text{TeV}}$ \\
\midrule
$C_{8g}^{}$               &                    $-105.5$ &   $-0.35 \pm 0.12$ &                            18 \\
$C_{SLL}^{s}$             &                      15.3 &    $0.05 \pm 0.02$ &                             7 \\
$C_{TLL}^{s}$             &                     $-43.9$ &   $-0.14 \pm 0.05$ &                            12 \\
$C_{SLL}^{c}$             &                     $-79.8$ &   $-0.26 \pm 0.09$ &                            16 \\
$C_{TLL}^{c}$             &                     $-46.7$ &   $-0.15 \pm 0.05$ &                            12 \\
$\widetilde{C}_{SLL}^{c}$ &                     $-68.5$ &   $-0.23 \pm 0.07$ &                            15 \\
$\widetilde{C}_{TLL}^{c}$ &                   $-1776.2$ &   $-5.9 \pm 1.9$ &                            76 \\
$C_{SLL}^{b}$             &                    $-105.6$ &   $-0.35 \pm 0.12$ &                            18 \\
$C_{TLL}^{b}$             &                     $-32.2$ &   $-0.11 \pm 0.03$ &                            10 \\
$\widetilde{C}_{SLL}^{b}$ &                    $-103.1$ &   $-0.34 \pm 0.11$ &                            18 \\
$\widetilde{C}_{TLL}^{b}$ &                   $-4070.4$ &  $-13.4 \pm 4.5$ &                           115 \\
\bottomrule
\end{tabular}

%% file: metab_C_all.tex
\begin{tabular}{lrrrrrrrcr}
\toprule
{} &  $\langle Q_g^- \rangle_0$ &  $\langle Q_1^{\text{SLL},u} \rangle_0$ &  $\langle Q_2^{\text{SLL},u} \rangle_0$ &  $\langle Q_3^{\text{SLL},u} \rangle_0$ &  $\langle Q_4^{\text{SLL},u} \rangle_0$ &  $\langle Q_1^{\text{SLL},d} \rangle_2$ &  $\langle Q_2^{\text{SLL},d} \rangle_2$ &               $P_i$ &  $\frac{\Lambda}{\text{TeV}}$ \\
\midrule
$C_{SLL}^{u}$             &                       $-0.2$ &                                    14.3 &                                   206.4 &                                   $-13.9$ &                                     4.5 &                                  $-119.0$ &                                  2541.5 &     $74 \pm 16$ &                           272 \\
$C_{TLL}^{u}$             &                       $-0.1$ &                                   163.3 &                                   $-50.3$ &                                    $-7.9$ &                                   $-36.6$ &                                 $-3625.2$ &                                 $-5846.8$ &   $-162 \pm 36$ &                           402 \\
$\widetilde{C}_{SLL}^{u}$ &                       $-0.1$ &                                    62.8 &                                    62.0 &                                   $-10.8$ &                                     0.5 &                                   534.9 &                                  $-496.7$ &     $-15.6 \pm 3.3$ &                           124 \\
$\widetilde{C}_{TLL}^{u}$ &                       $-3.5$ &                                   350.0 &                                   475.8 &                                  $-176.7$ &                                    38.6 &                                 $-1075.2$ &                                $-17591.7$ &  $-509 \pm 111$ &                           713 \\
\bottomrule
\end{tabular}

%% file: metab_D_all.tex
\begin{tabular}{lrrrrrcr}
\toprule
{} &  $\langle Q_g^- \rangle_0$ &  $\langle Q_1^{\text{SLL},d} \rangle_0$ &  $\langle Q_1^{\text{SLL},d} \rangle_2$ &  $\langle Q_2^{\text{SLL},d} \rangle_0$ &  $\langle Q_2^{\text{SLL},d} \rangle_2$ &             $P_i$ &  $\frac{\Lambda}{\text{TeV}}$ \\
\midrule
$C_{SLL}^{d}$ &                        0.8 &                                    $-6.1$ &                                   137.4 &                                   111.1 &                                 $-2493.8$ &  $-87 \pm 16$ &                           295 \\
$C_{TLL}^{d}$ &                       $-2.2$ &                                  $-162.6$ &                                  3649.5 &                                  $-254.3$ &                                  5708.9 &  $191 \pm 37$ &                           436 \\
\bottomrule
\end{tabular}

%% file: metab_E_all.tex
\begin{tabular}{llrlrcr}
\toprule
{} & $\langle Q_7 \rangle_2$ &  $\langle Q_8 \rangle_2$ & $\langle Q_1^{\text{SLR},u} \rangle_0$ &  $\langle Q_2^{\text{SLR},u} \rangle_0$ &              $P_i$ &  $\frac{\Lambda}{\text{TeV}}$ \\
\midrule
$C_{SLR}^{u}$             &                         &                   $-350.6$ &                                        &                                   187.4 &  $-266 \pm 20$ &                           515 \\
$\widetilde{C}_{SLR}^{u}$ &                    84.1 &                    $-88.8$ &                                     45 &                                    47.5 &    $-60 \pm 5$ &                           244 \\
\bottomrule
\end{tabular}